\documentclass[fleqn,usenatbib]{mnras}

\usepackage{newtxtext,newtxmath}
\usepackage[T1]{fontenc}
\usepackage{ae,aecompl}

\usepackage{graphicx}	% Including figure files
\usepackage{amsmath}	% Advanced maths commands
\usepackage{amssymb}	% Extra maths symbols
\usepackage{float}
\usepackage{multirow}
\usepackage{tikz}
\usepackage{enumitem}
\usepackage{xcolor}
\usepackage{ulem}
\usetikzlibrary{shapes,arrows}
\tikzstyle{decision} = [diamond, draw, fill=blue!30, 
    text width=7.0em, text centered, node distance=3cm, inner sep=0pt]
\tikzstyle{block} = [rectangle, draw, fill=orange!40, 
    text width=9.0em, text centered, rounded corners, minimum height=4em]
\tikzstyle{bigq} = [rectangle, draw, fill=yellow!40, 
    text width=9.0em, text centered, rounded corners, minimum height=6em]
\tikzstyle{cloud} = [draw, circle ,fill=red!40, text width=4.5em, text centered]
\tikzstyle{flagonoff} = [draw, circle ,fill=green!40, text width=4.5em, text centered]
\tikzstyle{arrow} = [thick,->,>=stealth]
\newcommand{\arepo}{\textsc{AREPO}}
\newcommand{\ie}{i.e.}
\newcommand{\eg}{e.g.}
\title[Resolving shocks and filaments in the CGM]{Resolving shocks and filaments in galaxy formation simulations: effects on gas properties and star formation in the circumgalactic medium}

\author[Bennett \& Sijacki]{
Jake S. Bennett,$^{1, 2}$\thanks{E-mail: jake.bennett@ast.cam.ac.uk}
Debora Sijacki$^{1, 2}$
\\
$^{1}$Institute of Astronomy, University of Cambridge, Madingley Road, Cambridge, CB3 0HA, UK\\
$^{2}$Kavli Institute for Cosmology Cambridge, University of Cambridge, Madingley Road, Cambridge, CB3 0HA, UK
}

\pubyear{2020}

\begin{document}
\label{firstpage}
\pagerange{\pageref{firstpage}--\pageref{lastpage}}
\maketitle

\begin{abstract}
There is an emerging consensus that large amounts of gas do not shock heat in the circumgalactic medium (CGM) of massive galaxies, but instead pierce deep into haloes from the cosmic web via filaments. To better resolve this process numerically, we have developed a novel `shock refinement' scheme within the moving mesh code \arepo\ that adaptively improves resolution around shocks on-the-fly in galaxy formation simulations. We apply this to a massive $\sim10^{12}$ M$_\odot$ halo at $z=6$ using the successful FABLE model, increasing the mass resolution by a factor of $512$. With better refinement there are significantly more dense, metal-poor and fast-moving filaments and clumps flowing into the halo, leading to a more multiphase CGM. We find a $\sim50$ per cent boost in cool-dense gas mass and a $25$ per cent increase in inflowing mass flux. Better resolved accretion shocks cause turbulence to increase dramatically, leading to a doubling in the halo's non-thermal pressure support. Despite much higher thermalisation at shocks with higher resolution, increased cooling rates suppress the thermal energy of the halo. In contrast, the faster and denser filaments cause a significant jump in the bulk kinetic energy of cool-dense gas, while in the hot phase turbulent energy increases by up to $\sim150$ per cent. Moreover, H\textsc{i} covering fractions within the CGM increase by up to $60$ per cent. Consequently star formation is spread more widely and we predict a population of metal-poor stars forming within primordial filaments that deep \textit{JWST} observations may be able to probe.
\end{abstract}

\begin{keywords}
hydrodynamics -- methods: numerical -- galaxies: formation -- galaxies: evolution -- galaxies: haloes -- intergalactic medium
\end{keywords}

%%%%%%%%%%%%%%%%%%%%%%%%%%%%%%%%%%%%%%%%%%%%%%%%%%

%%%%%%%%%%%%%%%%% BODY OF PAPER %%%%%%%%%%%%%%%%%%

\section{Introduction}

During the hierarchical gravitational collapse of primordial overdensities, haloes grow by accreting dark matter and baryons from the intergalactic medium (IGM). This baryonic component is what then feeds the formation of a galaxy at a halo's centre. The mechanism by which this feeding occurs is not fully understood, as it involves a complex interplay between accretion and merging dynamics, cooling, shock heating, and feedback from a variety of physical processes. All of these processes act in concert in the circumgalactic medium (CGM), which plays host to the baryon cycle that regulates the feeding of galaxies. Studying the CGM can therefore give us powerful insight into galaxy formation and evolution. 

In the `classical' model of accretion onto dark matter haloes, the way gas reaches the central galaxy depends on how quickly it can cool. If cooling times are long compared to dynamical times, gas shock-heats to the virial temperature of the halo as it infalls. It then reaches approximate virial equilibrium as it forms a hot atmosphere, which will then slowly cool onto the central galaxy -- `hot mode' accretion. If instead the ratio of cooling and dynamical times is small, a stable virial shock does not form and gas rapidly cools and accretes onto the central object -- `cold mode' accretion \citep{ClassicCGM1,ClassicCGM2,ClassicCGM3,ClassicCGM4}. Hot mode accretion, implemented into semi-analytic models \citep[\eg][]{HotAccretionSAM1,HotAccretionSAM2,HotAccretionSAM3}, reasonably reproduced observed galaxy masses, hence its early popularity. The formation of a stable accretion shock, whereby hot mode accretion dominates, has been shown analytically to apply only to high mass haloes with $M_\mathrm{halo} \gtrsim 10^{12} \mathrm{M}_\odot$ \citep{VirialShockThresh}, though recent work has suggested haloes accreting at less than a critical value can also form a hot atmosphere at significantly smaller halo masses \citep{Fielding2017,Stern2020}. 

In recent years, a significant amount of observational evidence has been found that points to a more complex, multiphase CGM across a wide variety of galaxy masses, including in massive haloes that should be able to form a stable virial shock and a hot atmosphere \citep{ObsMultiCGM0,ObsMultiCGM1,ObsMultiCGM2,ObsMultiCGM3}. Observations of high star formation rates in large galaxies at high redshift also imply a significant source of cold gas accretion \citep{HighzHighSF1,HighzHighSF2}. 

Hydrodynamical simulations have shown that in massive haloes, cold gas can bypass virial shocks and penetrate deep into the hot atmosphere via filaments \citep[see \eg][]{CoolMode1,CoolMode2,CoolMode3,ColdFilamentInflow1,CoolMode4,ColdFilamentInflow2}. However, it is notoriously difficult to model this multiphase gas that constitutes the CGM numerically. With all standard cosmological simulation techniques resolution is focused on the densest regions, namely the interstellar medium (ISM), so to increase global resolution to better resolve the CGM would rapidly make large simulations prohibitively expensive. There is a significant body of work that has analytically and numerically investigated idealised representations of the CGM, or simulations of small patches of a halo, to determine how a multiphase CGM may arise and survive \citep[see \eg][]{McCourt2012,McCourt2015,Scannapieco2015,Mandelker2016,Mandelker2020a,Lochhaas2020}, however these are not always set within the full galaxy formation framework that cosmological simulations provide and thus lack self-consistent treatment of cosmic gas inflows and outflows.   

To address the shortcomings of resolution in standard cosmological simulations, several recent works have developed novel refinement methods or globally increased the resolution in small cosmological boxes to investigate the CGM in more detail \citep[][see also \citealt{Vazza2010,MiniatiMatryoshka} for studies of how resolution affects turbulence]{HummelsCGM,Mandelker2019,PeeplesCGM,SureshCGM,RomulusC,vdVoortCGM,Nelson2020}. The criteria by which resolution in each of their simulations is increased varies, though most of them effectively refine uniformly within a preset radius around a halo of interest. IllustrisTNG50 is the only simulation in which the base resolution of the \textit{entire} simulation box is significantly boosted, though this comes at a large computational cost \citep[][see also a recent simulation project, Extreme Horizon, \citealt{ExtremeHorizon}]{TNG50_1,TNG50_2,Nelson2020}. The effects of higher resolution and refinement schemes vary between these works too, with \citet{HummelsCGM}, \citet{Mandelker2019}, \citet{vdVoortCGM} and \citet{Nelson2020} finding increases in H\textsc{i} column density and the amount of cool gas within their haloes, whereas \citet{PeeplesCGM} and \citet{SureshCGM} find much less of an effect. Many factors are likely to affect this however, including the different codes and galaxy formation models employed \citep[as shown in][]{Nelson2013,Nelson2015}, the halo mass and redshift that is studied, as well as the starting resolution of the comparison simulations. 

In this paper we introduce a different, physically and numerically motivated technique, where in addition to Lagrangian refinement we boost resolution around structure formation shocks. In particular we aim to better resolve the accretion shocks at the boundary of the CGM and IGM in massive haloes. This is the region that will determine whether primordial gas filaments can penetrate into the hot halo, and how turbulent motions are generated in the wake of curved shocks.  As well as this, shocks in low resolution simulations are often numerically broadened over a significant region which can potentially lead to inaccurate post-shock gas properties, for example through in-shock cooling \citep{Creasey}. Moreover, the accretion of cold IGM gas onto filaments themselves is usually poorly resolved in cosmological simulations, despite their importance in feeding galaxies and as potential sites for star formation themselves \citep[see \eg][]{Mandelker2016, Mandelker2018}. The weak shocks at their edge however make them a useful target of our new scheme. This work aims to investigate the maximum effect that increasing resolution around shocks can have within a zoom-in simulation, therefore we also target these weak shocks in this work.

To study the impact of our new scheme on the CGM, we resimulate a very massive $\sim10^{12}$~M$_\odot$ galaxy cluster progenitor at $z=6$ with the successful physical model employed in the FABLE suite of simulations \citep{FABLE1, FABLE2, FABLE3}. We have selected such a rare overdensity as it should be both largely virialised \textit{and} surrounded by a significant amount of substructure, allowing us to investigate the complex interplay of hot virialised gas with significant cold inflows. As we are simulating such a large overdensity we choose to only evolve the halo until $z=6$ in this proof-of-concept study. We note that our scheme is very flexible and can be readily applied to simulations of a number of different numerical and physical setups. For example, the scheme has the potential to allow us to target resolution increases in the CGM of multiple objects, as opposed to just one using the standard zoom-in technique. It can also be combined with different tracer fields \citep[\eg][]{Genel2013} to specifically target shocks caused by feedback from supernovae or black holes or it can be restricted to the desired range of shock strengths. The full capability of the model will be explored in future work. Our main aim in this paper is to isolate the sole effect of resolution on gas in the CGM and hence we also run a basic simulation, with most galaxy formation physics switched off, to determine if the trends we find are dependent on the implemented physical model. 

We start our investigation from a base mass resolution typical of some large scale galaxy formation simulations like Horizon-AGN \citep{HorizonAGN}, Magneticum \citep{Magneticum1,Magneticum2,Magneticum3} and the TNG300 run of the IllustrisTNG project \citep{TNG1,TNG2,TNG3,TNG4,TNG5}, so we can explore the limitations of such models. Other modern simulations start with a base mass resolution a factor of $\sim10$ better than this such as the $100$ Mpc boxes of Illustris \citep{Illustris1,Illustris2,Illustris3}, IllustrisTNG and EAGLE \citep{EAGLE1,EAGLE2}, though in this work we present results reaching and surpassing this resolution (by up to a factor of $50$) with our new scheme and still find a number of important changes and non-convergence in a number of physical quantities. A small number of simulations of smaller boxes including IllustrisTNG50 \citep{TNG50_1,TNG50_2} have a base resolution higher than these, though as previously mentioned these simulations rapidly become very expensive. 

The structure of this paper is as follows. Section~\ref{Section:Methods} introduces the simulations in more detail, before describing the newly implemented `shock refinement' scheme in Section~\ref{Section:RefinementMethods}. We present simple 2D Sedov-Taylor blast wave tests to verify our scheme in Section~\ref{Section:Verify}. We analyse our results by first visually inspecting the simulated halo in Sections~\ref{Section:BasicEffects} and~\ref{Section:Vis}, including a novel 3D visualisation of halo accretion. We consider the multiphase nature of the CGM in Section~\ref{Section:ColdGas} and the effect on kinematics and turbulence in the halo in Section~\ref{Section:Turbulence}. We investigate how these effects change the energy budget of the halo in Section~\ref{Section:EnergyBudget}, before looking at the impact on H\textsc{i} covering fractions and star formation in Section~\ref{Section:StarFormation}. We then summarise our results and conclude in Section~\ref{Section:Conclusion}. 

\section{Methods} \label{Section:Methods}
\begin{figure*}
\centering
\begin{tikzpicture}[node distance = 1.5cm, auto]
    \node [block] (shockfind) {Run shock finder to identify all shocked cells};
    \node [bigq, below of=shockfind, yshift=-0.25cm] (checkflag) {Is cell flagged?};
    \node [decision, left of=checkflag, xshift=-1cm] (pickshocked) {Is cell within $R_\mathrm{ref}$?};
    \node [block, above of=shockfind, yshift=0.1cm] (refine) {(De)refine to keep all cells within a factor of 2 of target mass};
    \node [block, above of=refine, yshift=0.15cm, line width=1.5mm] (nextstep) {Move to next timestep};
    \node [block, below of=pickshocked, yshift=-1.5cm] (findneighbours) {Identify all cells within smoothing length of shocked cell};
    \node [cloud, left of=refine, xshift=-2.5cm] (defaultref) {Set target mass to $m_\mathrm{base}$};
    \node [cloud, right of=refine, xshift=2.5cm] (lowertarget) {Set target mass to $m_\mathrm{base} / \alpha$};
    \node [decision, left of=pickshocked, xshift=-1cm] (radialcut) {Is cell Mach number greater than $\mathcal{M}_\mathrm{thresh}$?};
    \node [decision, right of=checkflag, xshift=1cm] (checktimer) {Has time $t_\mathrm{ref}$ passed since flagging?};
    \node [cloud, below of=checktimer, yshift=-1.5cm] (unflag) {Unflag cell};
    \node [cloud, below of=checkflag, yshift=-1.5cm] (flag) {Flag cells, record time};
    
    \draw [arrow] (shockfind) -- (checkflag);
    \draw [arrow] (pickshocked) -- node[anchor=west] {no} (defaultref);
    \draw [arrow] (pickshocked) -- node[anchor=north] {yes} (radialcut);
    \draw [arrow] (radialcut) |- node[anchor=south] {no} (defaultref);
    \draw [arrow] (radialcut) |- node[anchor=north] {yes} (findneighbours);
    \draw [arrow] (lowertarget) -- (nextstep.east);
    \draw [arrow] (defaultref) -- (nextstep.west);
    \draw [arrow] (refine) -- (shockfind);    
    \draw [arrow] (findneighbours) -- (flag);
    \draw [arrow] (nextstep) -- (refine);
    \draw [arrow] (checkflag) -- node[anchor=north] {yes} (checktimer);
    \draw [arrow] (checkflag) -- node[anchor=north] {no} (pickshocked);   
    \draw [arrow] (checktimer) -- node[anchor=east] {yes} (unflag);
    \draw [arrow] (checktimer) -- node[anchor=east] {no} (lowertarget);  
    \draw [arrow] (flag) -- (checkflag);
    \draw [arrow] (unflag) -- (checkflag);
\end{tikzpicture}
\caption{Schematic diagram of the shock refinement scheme we have implemented, describing the process that occurs at each timestep. At the beginning of each timestep (outlined in bold), \arepo's basic refinement scheme splits and merges cells to keep all cells within a factor of 2 of their individual target mass. The shock finder is then used to find all shocked cells and the cell flags are checked. If the cell is not already flagged, the distance of the cell from the halo centre and the Mach number of it are found, and if it lies outside of $R_\mathrm{ref}$ or has a Mach number below $\mathcal{M}_\mathrm{thresh}$ the target mass is left at the base level. Otherwise cells are flagged and the time of refinement is recorded. For flagged cells the time since refinement is checked, and if this is less than $t_\mathrm{ref}$ the target mass is decreased to the desired level. This new target mass is then passed through to the next timestep when the process is repeated. We note that the 3 blue diamond panels can be readily changed or removed if needed, depending on the target of the scheme.}
\label{plt:ShockRefineFlowChart}
\end{figure*}
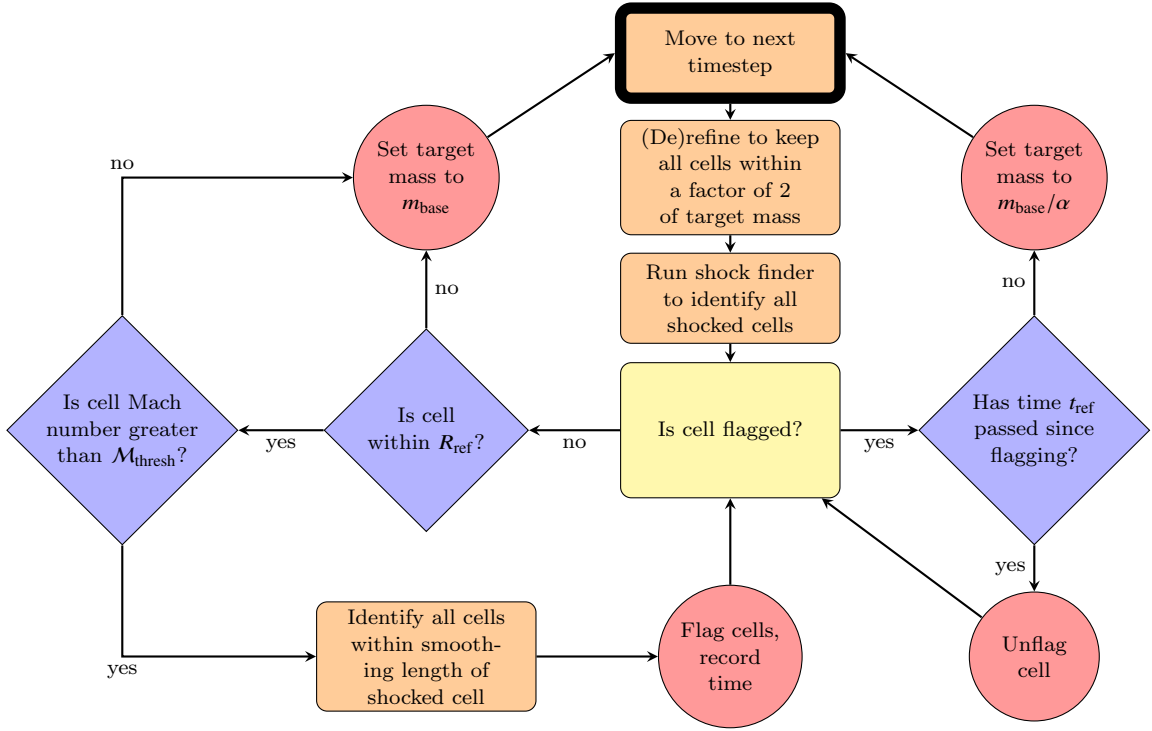  

\begin{figure} 
\raggedright \includegraphics[width=\linewidth]{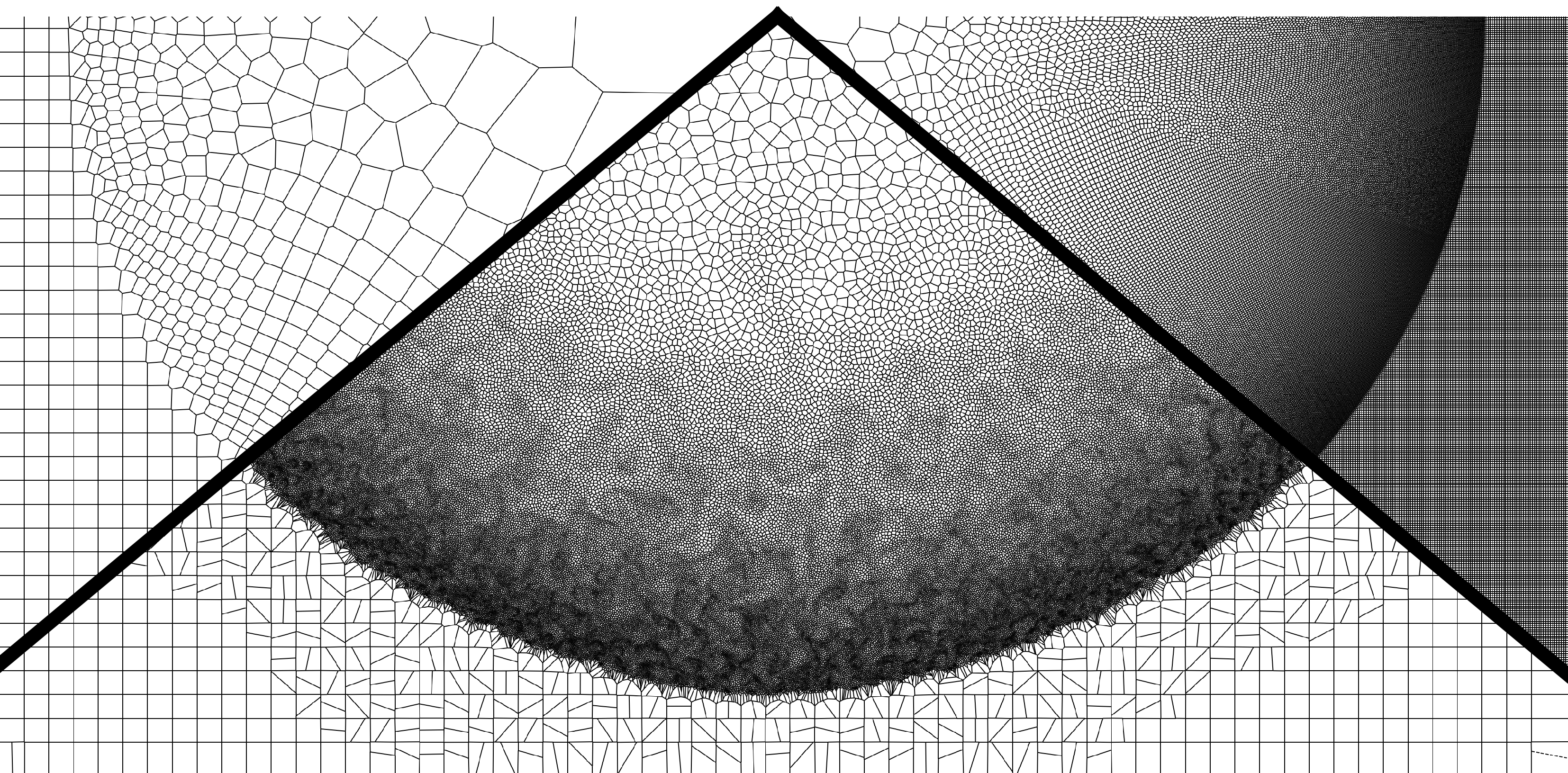}
\raggedright \includegraphics[width=\linewidth]{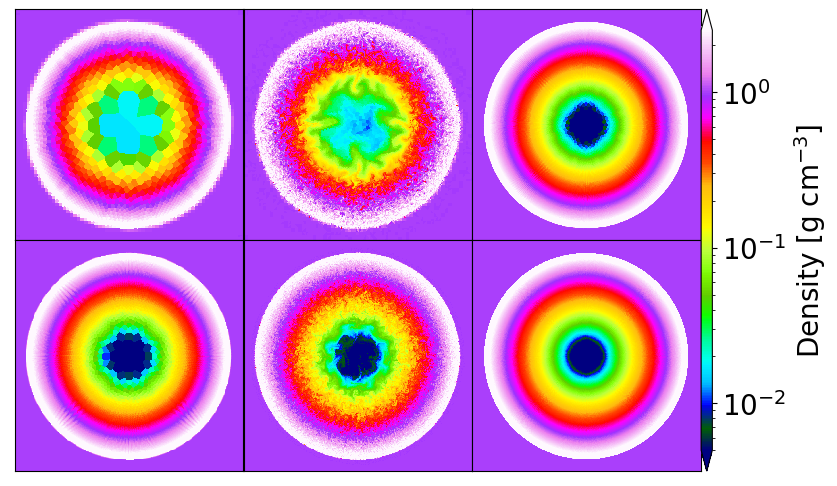}

\caption{\textit{Top panel}: Representation of the \arepo\ Voronoi mesh for the runs in the top row of the bottom panel. 
\textit{Bottom panel}: density maps of Sedov-Taylor blast simulations at time $t = 0.15$ for the following runs. \textit{Top row}: base refinement, $63^2$ grid (left); shock refinement, $63^2$ initial grid (centre); base refinement, $713^2$ grid (right). \textit{Bottom row}: base refinement, $255^2$ grid (left); shock refinement, $255^2$ initial grid (centre); base refinement, $2885^2$ grid (right). The right-hand column has a mass resolution a factor of 128 higher than the left hand column, which is equivalent to the \textit{maximum} level of refinement in the central column.}
\label{plt:SedovBlastImages}
\end{figure}

\begin{figure} 
\centering
\includegraphics[width=0.9\linewidth]{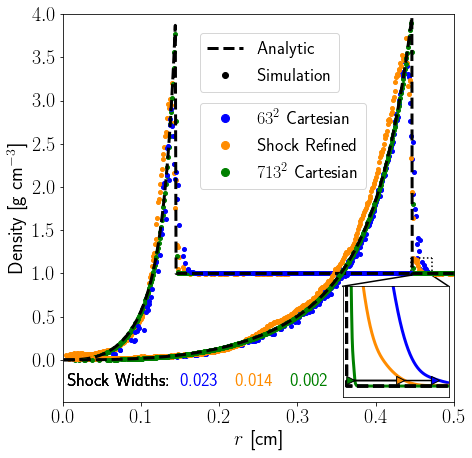}
\includegraphics[width=0.9\linewidth]{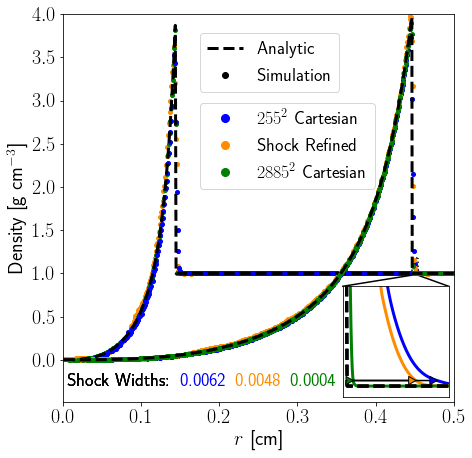}
\caption{Radial density profiles of the Sedov-Taylor blast wave at $t = 0.015$ and $0.15$. Binned simulation results are shown by the coloured points, and the analytic profile is illustrated by a black curve. \textit{Top panel}: runs starting from a $63^2$ Cartesian grid. \textit{Bottom panel}: runs starting from a $255^2$ grid. With shock refinement switched on, we note that the density peak is resolved with considerably more resolution elements and is sharper. This is demonstrated in the inset panels, where a zoomed in view of the pre-shock region is shown, with solid coloured lines showing cubic spline fits to the data. Coloured arrows show the radius at which the pre-shock density first rises above the value of $1.01$, which we define as the `shock width' and display the value of along the bottom of the plot. This shows how the numerical broadening of the shock is reduced when resolution is boosted.}
\label{plt:SedovBlastProfiles}
\end{figure}

\subsection{Simulation Code}
The simulations presented in this paper use the cosmological, hydrodynamical moving-mesh code \arepo\ \citep{Arepo}, which solves the Euler equations across a quasi-Lagrangian Voronoi mesh. We use cosmology consistent with Planck \citep{PlanckParameters}, where $\Omega_\mathrm{\Lambda} = 0.6911,$ $ \Omega_\mathrm{m} = 0.3089,$ $ \Omega_\mathrm{b}=0.0486,$ $\sigma_\mathrm{8} = 0.8159,$ $n_\mathrm{s} = 0.9667$ and $h=0.6774$. On-the-fly friends-of-friends \citep{FOF} and \textsc{Subfind} halo identification algorithms \citep{Subfind1,Subfind2} are used to identify gravitationally bound structures. 

We also make extensive use of the on-the-fly shock finder of \citet{ShockFinder}. This first identifies shock zones by finding cells meeting the following criteria:
\begin{align}
&\nabla \cdot \mathbf{v} < 0 \,,\\
&\nabla T \cdot \nabla \rho > 0 \,,\\
&\Delta \log T \geq \log \frac{T_2}{T_1}\Bigr|_{\substack{\mathcal{M}=\mathcal{M}_\mathrm{min}}} \wedge \ \ \ 
    \Delta \log p \geq \log \frac{p_2}{p_1}\Bigr|_{\substack{\mathcal{M}=\mathcal{M}_\mathrm{min}}} \,,
\end{align}
where $\mathbf{v}$, $T$, $\rho$, $p$ and $\mathcal{M}$ denote velocity, temperature, density, pressure and Mach number, respectively, and subscripts 1 and 2 denote pre-shock and post-shock states. If a cell is identified as being in a shock zone, a ray is sent from the centre of the cell along the temperature gradient. This is used to find the cells with the greatest compression and their corresponding Mach numbers by finding the pre- and post-shock temperatures along the ray. $\mathcal{M}_\mathrm{min}$ is a minimum Mach number that the shock finder reliably identifies, given as $\mathcal{M}_\mathrm{min} = 1.3$ in \citet{ShockFinder}. In our results any shocked cells with $\mathcal{M} < 1.3$ are therefore excluded as these shocks are potentially spurious. For a more extensive description of the shock finder see \citet{ShockFinder}.

\subsection{Physical Model} \label{Section:PhysicalModel}
We resimulate a massive galaxy cluster progenitor at high redshift using the zoom-in technique. This halo has previously been simulated in the FABLE suite described in \cite{FABLE1,FABLE2,FABLE3}, itself selected from a parent dark matter only simulation, Millennium XXL \citep{MilleniumXXL}. We choose a halo of mass $M_\mathrm{200} \approx 3.7 \times 10^{12}$ $\mathrm{M}_\mathrm{\odot}$ at $z=6$, henceforth referred to as the virial mass $M_\mathrm{vir}$\footnote{$M_\mathrm{200}$ and $M_\mathrm{500}$ refer to the mass contained within $R_\mathrm{200}$ and $R_\mathrm{500}$, respectively, within which the average density is $200$ and $500$ times the critical density of the Universe at the redshift of interest.}. This is a rare overdensity at this redshift, but it was chosen as we would expect a significant hot atmosphere as well as large amounts of cool, inflowing material. We can therefore investigate the interplay between cool filaments and a hot halo. The zoom-in has a region containing high resolution, collisionless dark matter particles, extending to approximately $10$~cMpc, each with a mass of $m_\mathrm{DM} = 5.54 \times 10^7 h^{-1} \mathrm{M}_\mathrm{\odot}$. Baryonic cell masses depend on the refinement scheme used as described in Section~\ref{Section:RefinementMethods}. 

We run simulations using two different sets of physical models. The first is an extremely basic scenario, henceforth referred to as `Reference' runs, where all explicit forms of feedback are switched off. We only include primordial cooling from H and He \citep{PrimordialCooling}, and switch off the uniform UV background. The formation of star particles is disabled, however we do allow ISM gas above a given density to become `star-forming'; \ie\ to move onto an effective equation of state (eEoS) that includes a pressure contribution from unresolved supernovae feedback \citep{eEoS_SFR}. This star formation number density threshold (henceforth referred to as the SFT) has a value of $n_\mathrm{H} = 0.2715$ cm$^{-3}$, which is the same as in FABLE. We also perform simulations with Reference physics and a boosted SFT of $n_\mathrm{H} = 100$ $\mathrm{cm}^{-3}$ to investigate the effect of refining very dense gas on our results. This is because the shock refinement does not operate in eEoS gas, of which there is a significant amount in the Reference runs due to catastrophic overcooling. The motivation behind all of the Reference runs is to clearly separate effects caused by changing numerical resolution from the effect of our chosen galaxy formation model. 

For the main body of results we run simulations with feedback enabled, using a slightly modified version of the sub-grid models described in Section~$2$ of \citet{FABLE1}, themselves a modified version of the Illustris feedback model \citep{Illustris1,  Illustris2, Illustris3}. In this paper we only include a single, central black hole in our halo of interest, which is seeded slightly later than its equivalent in FABLE. This black hole can also act as a tracer of the halo's potential minimum when the new shock refinement scheme is activated \citep[as used in][]{SureshCGM}. When the shock refinement scheme is active we aim to ensure that the stellar and active galactic nuclei (AGN) feedback of the FABLE model operate with a similar strength independent of resolution, so we can isolate effects due to changing refinement alone. We achieve this by verifying that the redshift evolution of the stellar and black hole masses remain very similar for all runs performed (for more details see Section~\ref{Section:RefinementMethods}). A final difference we have from FABLE is that the UV background of \citet{FG_UVB} is active from around $z\sim10.6$, rather than activated instantaneously at $z=6$ like in FABLE and Illustris. All other parameters are kept the same as in FABLE.

\subsection{Refinement} \label{Section:RefinementMethods}

In \arepo, each gas cell is given a target cell mass at the beginning of the simulation. \arepo's default refinement scheme then keeps all cells within a factor of two of this target mass by splitting and merging cells when needed (henceforth referred to as `base' refinement). A particularly useful characteristic of \arepo\ is that each cell within the simulation can have a different target mass, set according to almost arbitrary criteria. This allows numerical resolution to be focused on areas of interest outside of the densest regions without having to increase the global resolution of the simulation, which can be prohibitively expensive. This has been utilised in a number of different simulations to date, particularly within the context of AGN feedback \citep{CurtisRefinement, WeinbergerJet, BourneJet}.

In this work we have used this characteristic to develop a novel refinement technique to increase resolution on-the-fly around shocks, which are of fundamental importance in galaxy formation and evolution. In this `shock refinement' method, when all cells are active the on-the-fly shock finder of \cite{ShockFinder} is used to identify shocked cells. Cells with a Mach number above a threshold $\mathcal{M}_\mathrm{thresh}$ are then flagged for refinement, along with all cells within the shocked cell's smoothing length \citep[defined as twice the maximum radius of all Delaunay tetrahedra that have the cell at a vertex;][]{Arepo}. Flagged cells are given a lower target mass, $m_\mathrm{base} / \alpha$, where $\alpha > 1$ is a refinement factor. For our cosmological zoom-in simulations, the base resolution (BaseRef) has a target cell mass $m_\mathrm{base} = 1.64 \times 10^{7}$~M$_\odot$. We present results where $\alpha = 128$ (ShockRef128) and $512$ (ShockRef512), corresponding to a target cell mass of $m_\mathrm{ref} = 1.28 \times 10^5$~M$_\odot$ and $m_\mathrm{ref} = 3.2 \times 10^4$~M$_\odot$, respectively. 

In the runs presented in this paper we use a value of $\mathcal{M}_\mathrm{thresh} = 1.5$, which was chosen to refine around weak shocks, for example where gas accretes onto filaments, as well as strong virial shocks. This is a slightly more conservative limit than the limit $\mathcal{M}_\mathrm{min} = 1.3$ that the shock finder uses, so that we can be confident that we pick out genuine weak shocks. Due to the prevalence of shocks during galaxy formation (especially at high redshift) this has the effect of increasing resolution broadly throughout the simulation domain, though we note this is a deliberate choice for this proof-of-concept study. In future work this threshold can be adjusted or combined with other properties to target more specific regions of the simulation. To avoid refining around regions far outside the halo of interest we also impose an additional radial cut on the resolution criterion. We note that this was chosen as we are studying the virialised halo rather than the IGM in this work, and it would not be necessary in future work. To do this we track the centre of the halo by using either a passive particle (in the Reference runs) or the central black hole (in the FABLE physics runs) that moves with the halo's potential minimum. We set the cut off radius, $R_\mathrm{ref} = 214$ pkpc at $z=6$, following the definition of the IGM radius in \citet{SureshCGM}. 

All simulations are restarted from a preexisting run at $z=12$ with only the standard base refinement scheme active. At $z\sim10$ the shock refinement scheme is activated, chosen to allow sufficient time for the effects of higher resolution to manifest themselves by $z=6$ while keeping the cost of the simulation low. As mentioned above we also need to ensure that the stellar and AGN feedback operate in the same way at different resolutions, so we can isolate the effects due to increased resolution alone. We achieve this by increasing the number of neighbouring cells that feedback energy and metals are injected by a factor of $128$ for both the ShockRef128 and ShockRef512 runs with respect to the BaseRef run. This leads to stellar masses (within the halo's virial radius at $z = 6$) of $6.15\times10^{10}\mathrm{M}_\mathrm{\odot}$, $7.42\times10^{10}\mathrm{M}_\mathrm{\odot}$ and $8.13\times10^{10}\mathrm{M}_\mathrm{\odot}$ and to black hole masses at $z = 6$ of $9.92\times10^7\mathrm{M}_\mathrm{\odot}$, $8.19\times10^7\mathrm{M}_\mathrm{\odot}$ and $9.03\times10^7\mathrm{M}_\mathrm{\odot}$ for the BaseRef, ShockRef128 and ShockRef512 runs, respectively. We further discuss the effects of the refinement scheme on the stellar mass in Section~\ref{Section:StarFormation}. 
  
In the simulations with shock refinement we also record the time that a cell is flagged, and after a time $t_\mathrm{ref}$ the refinement flag is switched off. We choose a refine time $t_\mathrm{ref} = 10$ Myr in our cosmological simulations, the motivations for which are twofold. The first reason is numerical, as the refinement mechanism within \arepo\ is not instantaneous. Cell masses decrease by a factor of approximately two each timestep, and $10$ Myr is approximately twice the longest timestep in the simulation at $z=6$. We therefore use this timer to ensure flagged cells have the chance to refine first before they can derefine again. The second motivation is physical; in our simulated halo at $z=6$ this is the time taken for an inflowing filament to traverse $\sim10$ per cent of $R_\mathrm{vir}$, meaning that filaments can pass through the accretion shock and enter the hot halo before they would be allowed to derefine. We note however that our results are robust to reasonable changes in $t_\mathrm{ref}$. In future work this timer could be modified further depending on the regions of interest, for example by resolving the Sedov-Taylor phase of a supernova blast wave before lowering resolution again. 

As we have discussed, our scheme is very flexible and can be readily applied to a wide variety of simulations. A different range of $\mathcal{M}_\mathrm{thresh}$ values could be used to target different regions of interest around different haloes without globally increasing resolution. For example, in large box simulations a higher $\mathcal{M}_\mathrm{thresh}$ could be chosen to better resolve strong accretion shocks only. Moreover, our method could be applied to small scale simulations of the interstellar medium or accretion flows around black holes to track shocks generated by supernovae blast waves or AGN-driven jets and outflows with much higher accuracy. 

The entire shock refinement algorithm during each timestep of the cosmological simulations is shown in a schematic diagram in Fig.~\ref{plt:ShockRefineFlowChart}.

\subsection{Verification -- Sedov-Taylor blast wave} \label{Section:Verify}

Before presenting the main results of our cosmological simulations, we want to verify that our new scheme works as expected. To do this, we employ a simple 2D Sedov-Taylor blast wave simulation. We set up an initial Cartesian mesh with a uniform density $\rho_0 = 1$ g cm$^{-3}$ and internal energy $u_0 = 2.1 \times 10^{-22}$ erg. We then inject an internal energy $E = 1$ erg into the central cell. The resulting Sedov-Taylor shock wave has a well known analytic solution, allowing us to verify our refinement scheme. In all shock refined blast waves we use a refinement factor $\alpha = 128$, and compare this to the output from a simulation using a Cartesian grid with $128$ times higher resolution. This therefore means that the \textit{global} mass resolution of this second Cartesian grid is the same as the \textit{peak} mass resolution of the shock refinement run. We also switch off the timer described in Section~\ref{Section:RefinementMethods} to explicitly follow the post-shock evolution of the gas.

The advantages of the new refinement scheme are visually shown in Fig.~\ref{plt:SedovBlastImages}. The top panel shows, from left to right, a representation of the \arepo\ Voronoi mesh for the blast wave at $t=0.15$, for a base refinement run with a $63^2$ Cartesian grid, the shock refinement run starting from a $63^2$ Cartesian grid and a base refinement run with a $713^2$ Cartesian grid. In the bottom panel of Fig.~\ref{plt:SedovBlastImages} we plot the density field at $t = 0.15$ for 6 runs: base refinement runs with initial Cartesian grids of $63^2$ and $255^2$ (top left and bottom left, respectively), shock refinement runs with the same initial Cartesian grids (top centre and bottom centre), and base refinement runs with a higher resolution initial Cartesian mesh, equivalent to \textit{globally} having the peak mass resolution of the shock refinement scheme ($713^2$, top right, and $2885^2$, bottom right). The shock front itself is much sharper and better defined in both of the higher resolution runs, with considerably more resolution elements in place in the high density shell. In the shock refinement run the low resolution of the undisturbed material is maintained, reducing computational expense compared to the high resolution Cartesian grid. Furthermore, cells already begin to refine ahead of the shock, due to the flagging of the neighbours of shocked cells, allowing us to better capture the progression of the blast wave than in the low-resolution simulation. 

As a by-product of cells splitting and merging triggered by shock refinement, small density inhomogeneities are created in shock refinement runs that then seed Rayleigh-Taylor instabilities in the post-shock gas. These inhomogeneities are not present in the base simulations due to the extreme symmetry of the initial set up, meaning that the growth of instabilities is largely suppressed. The Rayleigh-Taylor instabilities, and the associated fluid mixing, are more pronounced in the lower resolution run as the refinement of an initially lower resolution grid generates larger seed inhomogeneities. We note that supernova sites, which are prototypical examples of this kind of blast wave, are embedded within an inhomogeneous, multiphase ISM and are well known to exhibit Rayleigh-Taylor instabilities. However, matching the observed properties and the time evolution of these instabilities is beyond the scope of our study. 

We now turn to a more quantitative analysis and compare the simulated density profiles with analytic results in Fig.~\ref{plt:SedovBlastProfiles}. The shock refinement runs (orange) are able to more accurately capture the analytic density peak, particularly at early times, than the low resolution runs they are based on (blue). This is highlighted within the inset panels of Fig.~\ref{plt:SedovBlastProfiles}, where we show a cubic spline fit to the data in the pre-shock region near to the discontinuity. We then define a `shock width' as a distance from the analytic shock front to the point at which this fit in gas density first rises above a value of $1.01$. This shows that the numerical broadening of the shock front is reduced when our scheme is active. The level of improvement is better when starting from a lower initial resolution, where the shock is $\sim 40$ per cent narrower compared to $\sim 23$ per cent for the run with higher initial resolution. In terms of run time shock refinement is more competitive against the high resolution Cartesian runs (green), albeit with considerably more scatter in the post-shock region stemming from the fluid instabilities discussed above. This leads to a slight offset between the shock refinement and analytic density profiles, especially at low resolution, given that the analytical treatment does not include any instabilities. 

We conclude from this that the shock refinement scheme can significantly improve the capture of simulated blast waves, with numerical broadening reduced the most when starting from low-resolution initial conditions even if larger instabilities cause more deviation from the simple analytic expectation downstream.

\section{Results} \label{Section:Result}

\begin{table*}
	\centering
	\caption{Summary of the parameters employed in each simulation run. Columns two to four list the refinement scheme employed, density threshold for star formation and target gas mass, respectively, while the remaining columns list the number of cells and the values of the 10th and 90th percentile cell masses and radii for all cells within $R_\mathrm{ref}$.}
	\label{tab:runtable}
	\begin{tabular}{c|c|c|c|c|c|c|c|}
        \hline
        Physical Model & Refinement & SFT [cm$^{-3}$] &  Target Mass [M$_\odot$] & No. of Cells &  Cell Mass Range [M$_\odot$] & Cell Radii Range [kpc] \\
        \hline
        \hline

        Reference & BaseRef & $0.2715$ & $1.64\times 10^7$ & 99,562 & $1.01\times 10^7 - 2.40\times 10^7$ & $0.05 - 6.31$\\
        (only & ShockRef128 & $0.2715$ & $1.28\times10^5$ & 4,326,240 & $1.02\times 10^5 - 2.85\times 10^5$ & $0.23 - 1.57$ \\
        primordial & BaseRef & $100$ & $1.64\times 10^7$ & 89,283 & $1.07\times 10^7 - 2.68\times 10^7$ & $0.04 - 6.65$\\
        cooling) & ShockRef128 & $100$ & $1.28\times 10^5$ & 7,297,829 & $9.86\times 10^4 - 2.21\times 10^5$ & $0.02 - 1.14$\\ 
        \hline
        \multirow{3}{*}{FABLE model} & BaseRef & $0.2715$ & $1.64\times 10^7$ & 92,875 & $1.06\times 10^7 - 2.42\times 10^7$ & $0.72 - 6.31$\\
        & ShockRef128 & $0.2715$ & $1.28\times10^5$ & 6,943,981 & $1.04\times 10^5 - 2.43\times 10^5$ & $0.25 - 1.22$\\
        & ShockRef512 & $0.2715$ & $3.20\times 10^4$ & 26,019,084 & $2.57\times10^4 - 5.98\times10^4$ & $0.16 - 0.75$\\
        \hline
	\end{tabular}
\end{table*}

\begin{figure*}
\centering
\includegraphics[width=1\linewidth]{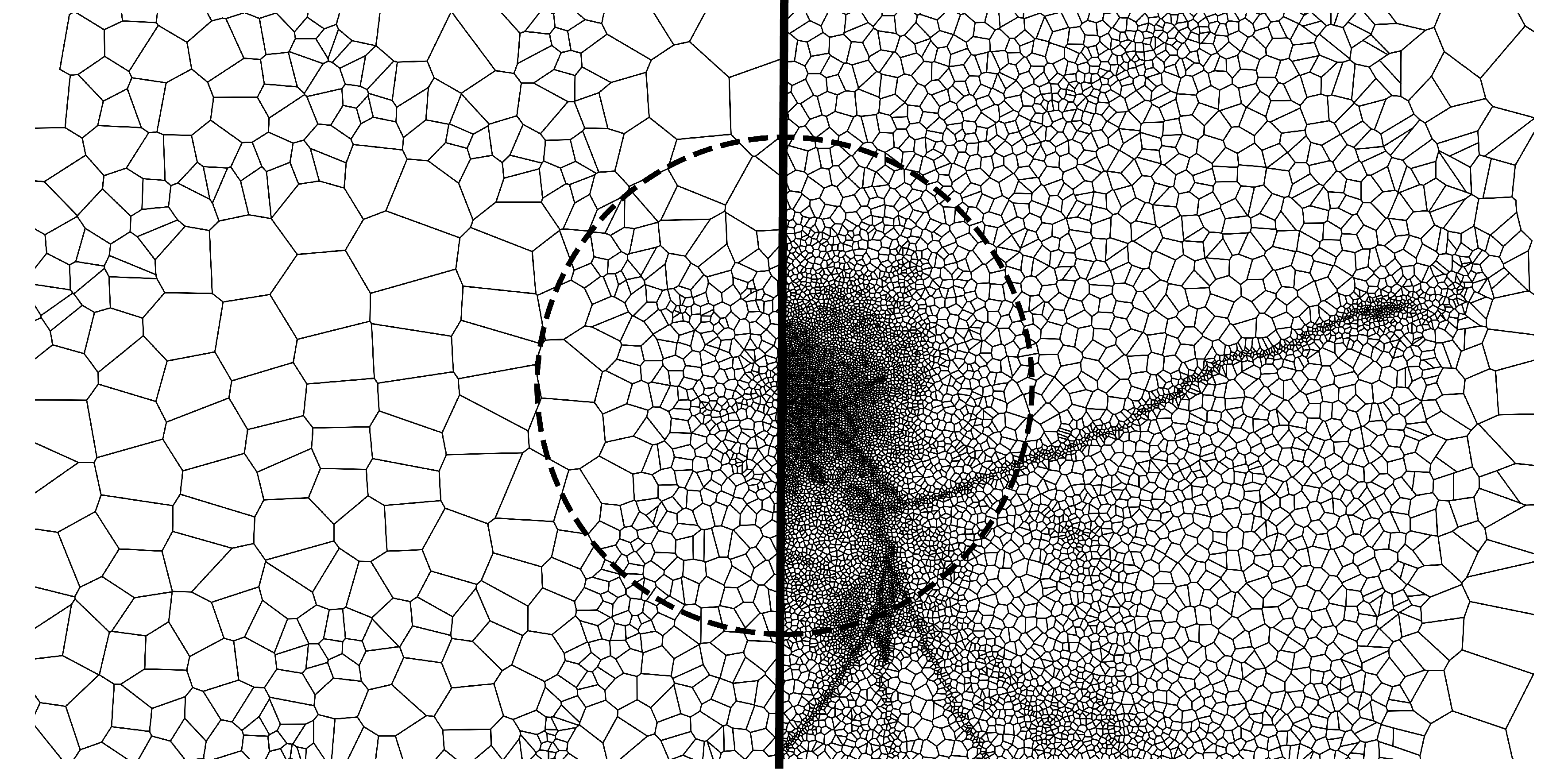}
\includegraphics[width=1\linewidth]{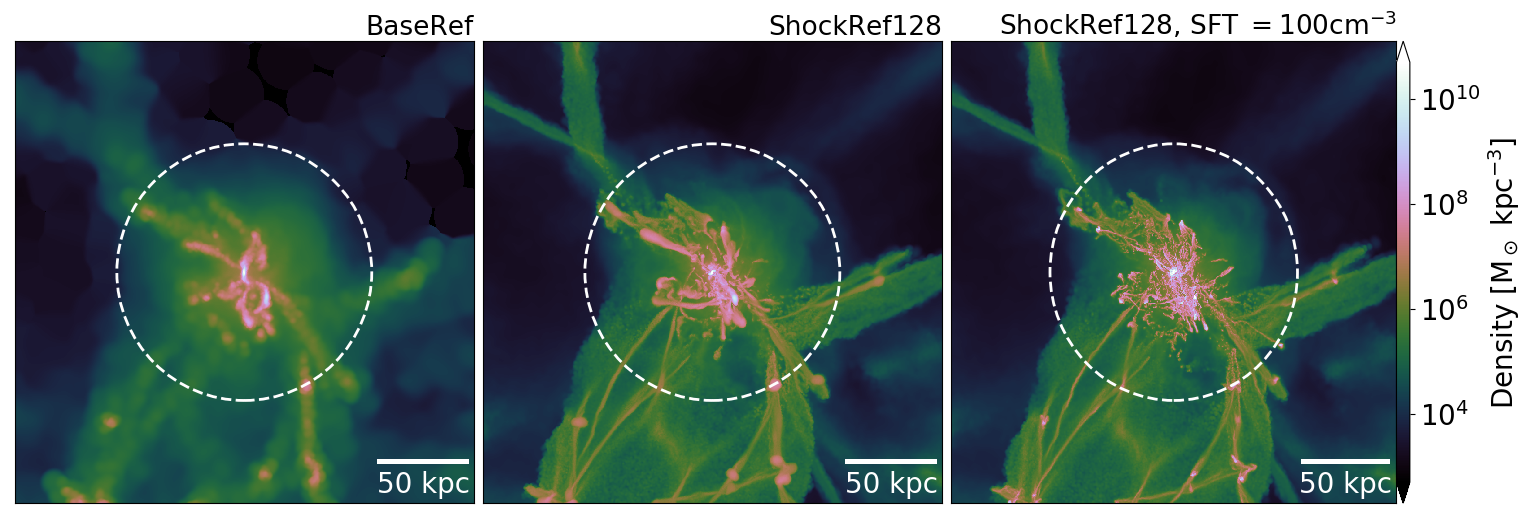}
\caption{\textit{Top panel}: Representation of the \arepo\ Voronoi mesh in the Reference physics run with base refinement and the SFT of FABLE (left) and the ShockRef128 run with a SFT of $n_\mathrm{H} = 100 \ \mathrm{cm}^{-3}$ (right). The number of resolution elements is hugely increased, and filamentary structure is particularly highlighted when the shock refinement scheme is active. \textit{Bottom panels}: Mass-weighted mean density maps of Reference runs within a slice of thickness $0.15 R_\mathrm{vir}$ around the halo centre. From left to right the runs are: BaseRef with SFT of FABLE; ShockRef128 with SFT of FABLE; ShockRef128 with SFT of $n_\mathrm{H} = 100 \ \mathrm{cm}^{-3}$. Dashed white circles indicate $R_\mathrm{vir}$.}
\label{plt:ClusterVoronoi}
\end{figure*}

\begin{figure*} 
\centering
\includegraphics[width=1\linewidth]{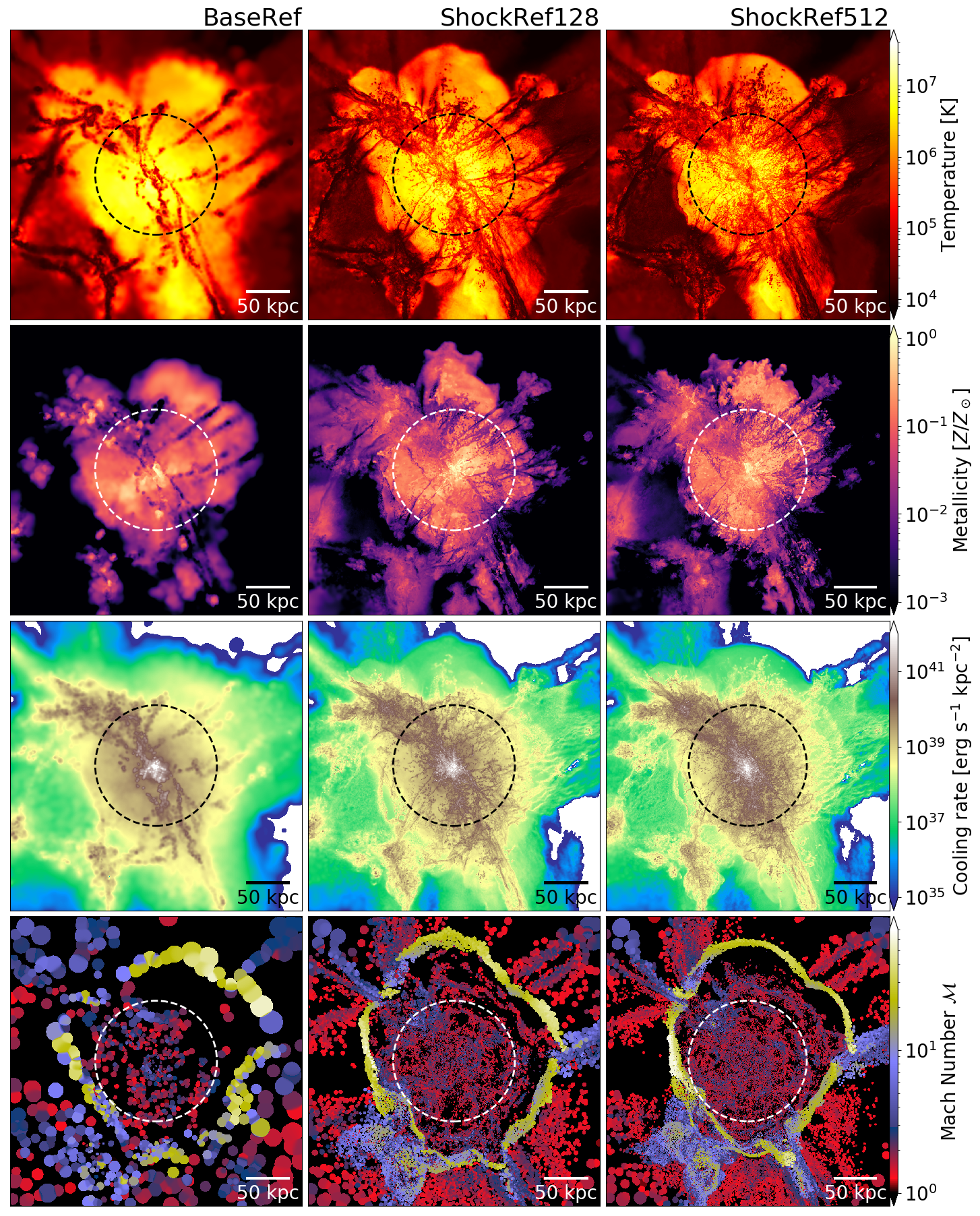}
\caption{\textit{From top to bottom}: maps of density-weighted mean temperature, density-weighted mean metallicity, cooling rate density, and energy dissipation-weighted mean Mach number projected over slices of thickness $1$, $1$, $1$ and $0.25$ $R_\mathrm{vir}$, respectively. The left-hand column shows the BaseRef run, the centre shows the ShockRef128 run, and the right-hand column the ShockRef512 run. The white areas in the cooling rate map correspond to regions being heated by the UV background. When the shock refinement scheme is active a much more multiphase CGM is developed, with cold, metal-poor filaments coexisting with a hot, metal-enriched gaseous halo.}
\label{plt:ClusterVis}
\end{figure*}

\begin{figure*} 
\centering
\includegraphics[width=1\linewidth]{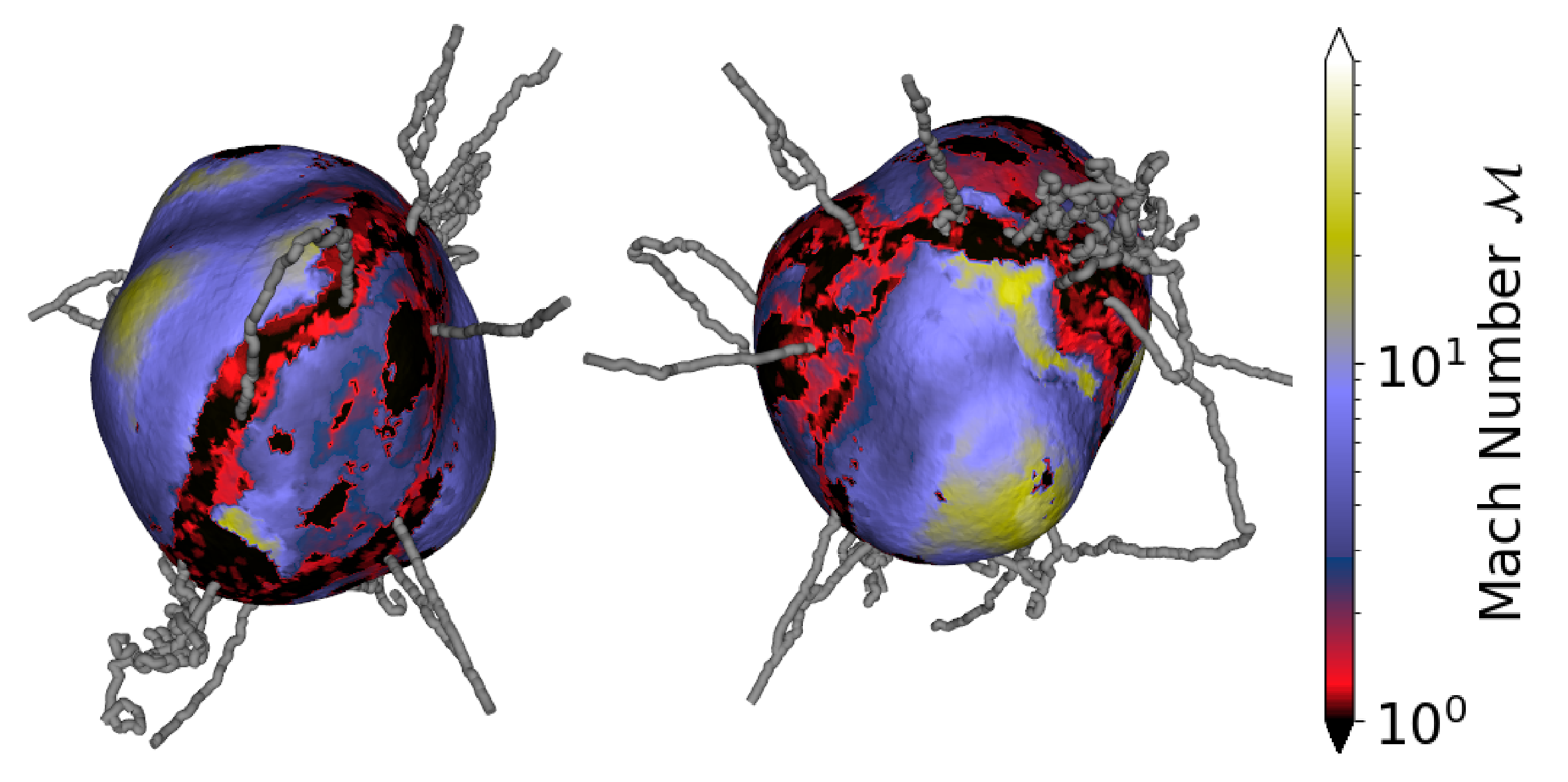}
\caption{3D mesh showing the accretion shock from two different orientations for the ShockRef128 run, with the surface coloured by the local Mach number. Yellow-white regions indicate where the accretion shock is strongest, corresponding to the source of hot halo gas. Black regions correspond to where there is a very weak accretion shock. The location of inflowing gas filaments found from DisPerSE \citep{disperse} are shown in grey. Their entry into the halo is clearly correlated with where the local Mach number is lowest. Visualisation created using \textsc{VTKplotter} \citep{vtkplotter} using the method described in the main text.}
\label{plt:AccretionShock3D}
\end{figure*}

We summarise the simulations we ran of our target $\sim10^{12}$ M$_\odot$ halo to $z=6$ in Table~\ref{tab:runtable}. The physical model used in each simulation is listed, along with the mass and spatial resolution of each run (considering the region within $R_\mathrm{ref}$). 

\subsection{Refinement Effects on Reference Runs} \label{Section:BasicEffects}

Fig.~\ref{plt:ClusterVoronoi} illustrates the impact of shock refinement on the Reference runs with a basic physical model. In the top panel we show a representation of \arepo's Voronoi mesh, with the BaseRef run on the left and the ShockRef128 with a SFT of $n_\mathrm{H} = 100$ cm$^{-3}$ on the right. Overall the shock refined run has many more resolution elements than the base run ($7.3$ million vs $100,000$ within $R_\mathrm{ref}$, as shown in Table~\ref{tab:runtable}). Perhaps the most striking difference in this image is the sharply resolved cosmic filaments, which are picked out as we refine around the weak shocks at their edge. There is also a significantly increased number of cells within the virial radius due to refinement around the accretion shock, which in the Reference runs without feedback lies mostly within $R_\mathrm{vir}$ at radii of between between $40$ and $70$~kpc. There is also much higher resolution in the centre of the halo for numerous reasons: the timer discussed in Section~\ref{Section:RefinementMethods} keeps cells refined for a $10$ Myr after passing through the accretion shock; numerous shocks are generated within the halo, including bow shocks from supersonic substructures and from supersonic filaments shocking as they deposit gas towards the inner regions of the halo. 
 
Maps of mass-weighted mean gas density for the Reference runs are shown in the bottom panel of Fig.~\ref{plt:ClusterVoronoi}, where we show the BaseRef and ShockRef128 runs with the same SFT as FABLE (left and centre, respectively) and the ShockRef128 run with a boosted SFT (right). The enhancement in the density of filaments and sheets in the two shock refinement runs is clear with respect to the base refinement. In the central and right-hand maps we compare a star formation density threshold of $n_\mathrm{H} = 0.2715$ cm$^{-3}$ and $n_\mathrm{H} = 100$ cm$^{-3}$, to gauge the impact of allowing very high density cells to refine. The density in filaments is smoother in the central image, due to substructures within the filaments being less well resolved. There is also a clear difference in the centre of the halo, where the density distribution is more complex and substructures are considerably smaller and denser in the right-hand panel. This lack of refinement in the unresolved ISM is a limitation of this model when using a relatively low SFT. In the Reference runs we are free to boost the SFT to reduce the amount of gas ending up on the eEoS. This has the adverse effect of increasing the run-time of the simulation, but allows us to determine cleanly the effect of allowing shock refinement to boost resolution in already dense gas\footnote{We also ran a base refinement simulation with a SFT of $n_\mathrm{H} = 100$ cm$^{-3}$ that shows qualitatively similar results, though we do not show this for brevity.}. In simulations with FABLE physics we cannot arbitrarily change the density threshold for star formation without drastically affecting the simulation's implementation of star formation and feedback, and hence FABLE's agreement with observations. We note however that as the runs in Fig.~\ref{plt:ClusterVoronoi} do not have feedback, gas cools, clumps and passes the SFT much more readily than in simulations including feedback. This can therefore be considered a worse case scenario. We also note that much of the inflowing filaments and more complex density structures in the wake of the accretion shock are largely unchanged by adjusting this parameter. 

\subsection{Gaseous Halo Properties with the FABLE model} \label{Section:Vis}

Now that we have inspected the basic changes the shock refinement scheme makes, we turn to examine its impact in more detail when implemented with a more realistic physical model. We reiterate that the FABLE physics runs use a fixed SFT of $n_\mathrm{H} = 0.2715$ cm$^{-3}$ and that the accretion shock is now located at considerably larger radii than in the Reference runs. 

Fig.~\ref{plt:ClusterVis} shows maps of density-weighted average temperature, density-weighted average metallicity, cooling rate surface density and energy dissipation-weighted Mach number for the FABLE physics runs contrasting the BaseRef run (left-hand column) to the ShockRef128 (centre column) and ShockRef512 (right-hand column) runs. In the temperature maps (top panels) it is evident that in both shock refinement runs structures of cool gas are much more prevalent inside the hot halo. Cool filaments pierce deep inside the virial radius, denoted by the black dashed circle, at all resolutions, but are more numerous, better defined and pierce deeper when shock refinement is active, leading to cool gas exhibiting a much more complex morphology. Furthermore, the boundary between the CGM and the IGM, at the edge of the gaseous halo, is noticeably sharper. Comparing the ShockRef128 and ShockRef512 runs we note that the multiphase nature of the CGM is very sensitive to the resolution, and that as we increase resolution further we find many more dense clumps and structures embedded within the hot halo gas \citep[for similar recent results see][]{Nelson2020}.  

Similar trends are seen in the metallicity maps (second row), where we find that the cool gas structures correspond primarily to low-metallicity filaments. This therefore points to an increase in largely primordial gas bypassing the accretion shock and streaming towards the central galaxy. In the metallicity maps the hot, enriched gas ejected by supernovae and AGN feedback is shown in yellow. These outflows are pushed out into the halo where they co-exist with cold, primordial inflows, indicating that with shock refinement we are capturing a much more multiphase CGM. The filaments in the high resolution run are particularly clear in the map of cooling rate density (third row), where the increase in dense gas at a temperature close to the peak of the hydrogen cooling curve leads to a boosted cooling rate throughout the CGM. This enhancement in cooling throughout the filaments and CGM will have clear implications for the energy budget of the halo, which we discuss further in Section~\ref{Section:EnergyBudget}.

The maps of Mach number (bottom row) show the edge of the CGM very clearly, with the high Mach number accretion shocks ($\mathcal{M} > 20$) highlighted in yellow-white at the boundary of the CGM and IGM. In the BaseRef run the accretion shock is not resolved as a coherent structure, whereas in both shock refined runs the shock becomes very well defined. The shape of the gaseous halo itself is modified when resolution is increased, stemming from the change in topology of the large-scale accretion shocks themselves with resolution. Although the most striking differences in the properties of the accretion shock are between the BaseRef and ShockRef128 runs, when the resolution is pushed further in the ShockRef512 run the shocks further narrow and somewhat change their location, strength and curvature. Recall that in the shock refinement scheme we increase resolution around all shocks above $\mathcal{M}_\mathrm{thresh} = 1.5$, corresponding to most of the cells visible in the bottom row of Fig.~\ref{plt:ClusterVis}. It is worth noting here how prevalent shocks are, meaning the shock refinement scheme is continually active in a large portion of the halo. Resolution is therefore widely increased throughout. Moreover, the scheme is effective at boosting resolution around the edge of filaments, which is demonstrated in the top-right of the shock refinement panels where the parallel lines of blue colour corresponds to an intermediate shock ($\mathcal{M}\sim3$) from IGM gas accreting onto either side of the inflowing filament. 

While the maps in Fig.~\ref{plt:ClusterVis} show the properties of the runs with FABLE physics activated, we find similar trends in our Reference runs, albeit with the accretion shock located at much smaller radii, indicating the qualitative behaviours we find here are independent of galaxy formation model. 

Having a well resolved accretion shock allows us to visualise it in three dimensions by creating an `accretion shock mesh' which can then be rendered. To do this we first identify cells associated with a strong shock along a number of sight-lines from the halo centre. The largest connected region of this point cloud is then found using a Friends-of-Friends algorithm which is then smoothed using a moving least squares algorithm within the 3D plotting package \textsc{VTKplotter} \citep{vtkplotter}. A surface reconstruction is then performed within \textsc{VTKplotter}, giving a 3D mesh representing the accretion shock. This initial mesh contains a number of holes where no strong accretion shock was found along the original sight-lines; to generate a closed surface we then add additional points at interpolated radii to fill the holes. We therefore end up with a closed 3D surface containing the extent of the halo's hot atmosphere, bounded by the accretion shock.

We show this surface for the ShockRef128 run with FABLE physics in Fig.~\ref{plt:AccretionShock3D}, where we display two different orientations of the halo. The surface is coloured according to the local Mach number. Interestingly, the accretion shock is clearly not spherical, with a shape dependent on the direction of accretion onto the halo. Some areas have a very high Mach number ($\mathcal{M} > 20$), indicating where the bulk of shock heating is occurring that generates the hot halo. However, a significant proportion of the surface is not associated with strong accretion shocks. In grey we also show gas filaments found with the DisPerSE algorithm \citep{disperse}, demonstrating that dense filaments of largely primordial gas penetrate the accretion shock where the local Mach number is lowest and deliver cold gas deep within the halo. There are many areas of the surface with low Mach number that do not have an associated filament, most likely corresponding to accretion from dense sheets that also remain largely unshocked, or from smaller filaments not found by DisPerSE. Fig.~\ref{plt:AccretionShock3D} gives us a clear visual representation of the different modes of gas accretion onto massive haloes, and indicates how neighbouring parts of the CGM can be in significantly different thermal states. This therefore shows how a multiphase CGM more in accord with observations can be, at least in part, generated from the topology of accretion onto massive haloes.

\subsection{Impact of Refinement on the Multiphase CGM} \label{Section:ColdGas}

\begin{figure} 
\centering
\includegraphics[width=\linewidth]{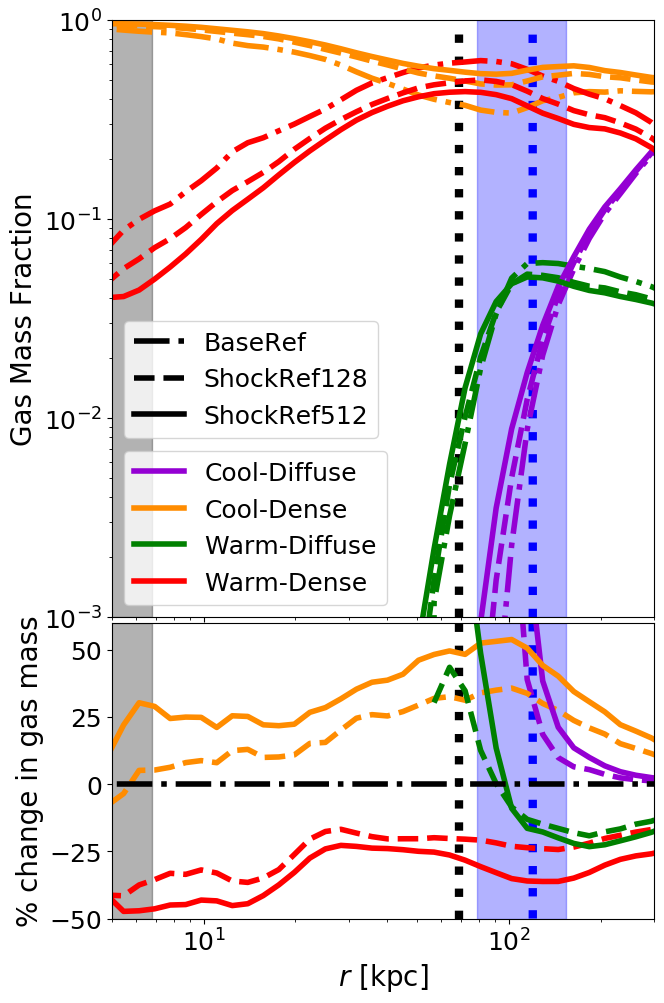}

\caption{\textit{Top panel}: Radial profiles of the fraction of cumulative gas mass in different phases relative to the total cumulative gas mass, for the three refinement runs. Cool gas is defined as having a temperature $T < 10^5$K, and dense gas as having a density $n_\mathrm{H} > 10^{-3} \mathrm{cm}^{-3}$. \textit{Bottom panel}: percentage change in cumulative gas mass in each phase as a function of radius, when the shock refinement scheme is switched on. For both panels, the black dotted vertical line represents $R_\mathrm{vir}$, and the shaded grey area indicates the region below $2.8$ times the gravitational softening length of the simulation. The blue dotted vertical line and associated shaded region indicate the median position and radial extent of the accretion shock in the ShockRef128 run, as described in the text.}
\label{plt:ColdGasFrac}
\end{figure}

After a visual inspection of our simulations in Section~\ref{Section:Vis}, we now turn to a more quantitative exploration of them. In the top panel of Fig.~\ref{plt:ColdGasFrac} we show radial profiles of the fraction of cumulative gas mass contained within cool ($T < 10^5$ K), warm ($T > 10^5$ K), diffuse ($n_\mathrm{H} < 10^{-3}$ $\mathrm{cm}^{-3}$) and dense ($n_\mathrm{H} > 10^{-3}$ $\mathrm{cm}^{-3}$) phases for the FABLE physics runs with the BaseRef, ShockRef128 and ShockRef512 runs. The black dotted line indicates $R_\mathrm{vir}$ and the blue dotted line and shaded region show the median radius and radial extent of the aspherical accretion shock in the ShockRef128 run, which we find from the radii of vertices of the accretion shock mesh shown in Fig.~\ref{plt:AccretionShock3D}. As suggested by the cooling rate maps in Fig.~\ref{plt:ClusterVis}, there is a boost in cool-dense gas (orange) throughout the CGM at the expense of the warm-dense phase (red). The cool-diffuse gas (shown in purple) that dominates the IGM does not exist interior to the accretion shock, though it persists down to smaller radii in the shock refinement runs due to the change in the shape and thickness of the accretion shock. The gas fraction of warm-dense gas (in red) peaks at the inner edge of the accretion shock, as the diffuse gas is compressed and heated. This peak is lower when shock refinement is active as more gas remains in the cool-dense phase, and there is also suppression of hot gas outside the accretion shock.

The percentage change in cumulative gas mass with respect to the BaseRef run within each gas phase is shown in the bottom panel of Fig.~\ref{plt:ColdGasFrac}. There is a significant boost in cool-dense gas mass within the CGM of $\sim30$ per cent for the ShockRef128 run and of $\sim50$ per cent for the ShockRef512 run. We note that the amount of mass in the cold-dense phase is converging weakly with resolution probed in our study, indicating that higher resolution simulations would predict an even more multiphase and clumpy CGM. Furthermore, there is a suppression of warm-dense gas within the CGM of $\sim20$ per cent and $\sim30$ per cent for the ShockRef128 and ShockRef512 runs, respectively. The qualitative changes we find here are encouragingly similar to other work \citep[\eg][]{HummelsCGM}, despite the different codes and drastically different redshift considered. Our results do deviate however from those of \citet{SureshCGM}, who find little change in cool-dense gas with boosted refinement, though their `base' resolution is a factor of two higher than our highest resolution run so this is perhaps not surprising. We reiterate that in this work we wish to investigate the differences that may be found compared to coarser large scale galaxy formation simulations such as Illustris or FABLE, which we find are significant.

The reduction in hot gas reduces the average temperature of the halo, which has implications for the thermal energy budget as discussed in Section~\ref{Section:EnergyBudget}. The significant boost of cold gas within the hot halo naturally creates a much more multiphase CGM in line with recent observations \citep[currently only available at lower redshifts, see \eg:][]{ObsMultiCGM0,ObsMultiCGM1,ObsMultiCGM2,ObsMultiCGM3}, and the presence of cold, dense clumps far from the central galaxy will also have obvious implications for the covering fraction of neutral hydrogen and for star formation, as we investigate in Section~\ref{Section:StarFormation}.     

\subsection{CGM Kinematics and Turbulence} \label{Section:Turbulence}
\begin{figure*} 
\centering
\includegraphics[width=\linewidth]{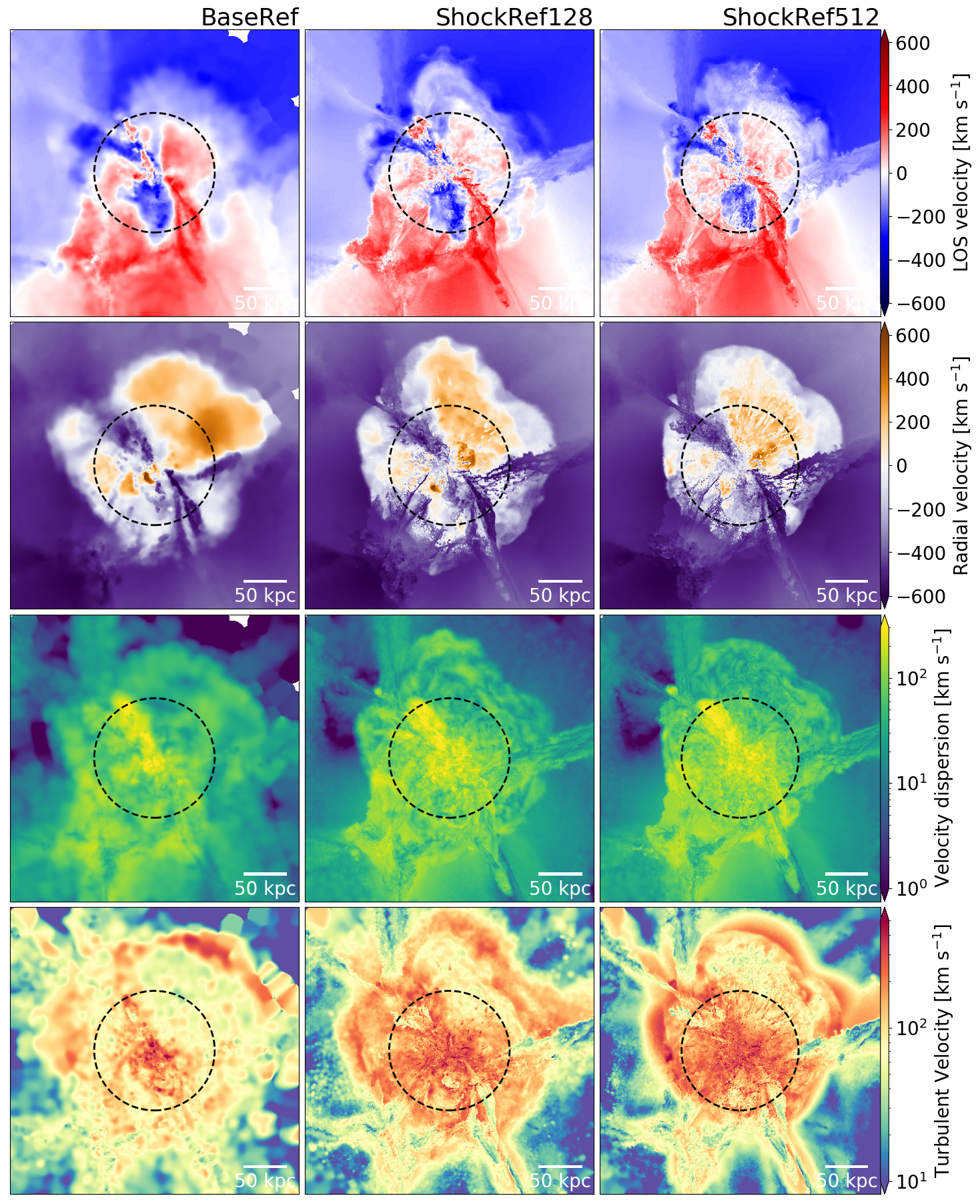}
\caption{\textit{From top to bottom}: maps of mass-weighted averages of LOS velocity, radial velocity, velocity dispersion along the LOS and turbulent velocity, all projected over a slice of thickness $0.25$ $R_\mathrm{vir}$. The left-hand column shows the BaseRef run, the central column shows the ShockRef128 run and the right-hand column shows the ShockRef512 run. Black dashed circles indicate $R_\mathrm{vir}$. The kinematic structure of the halo is much more complex when shock refinement is active. }
\label{plt:TurbMaps}
\end{figure*}

Given that increased resolution affects the thermal state of the gas in the halo and the propagation of filaments, changes in gas dynamics are to be expected. The top row of Fig.~\ref{plt:TurbMaps} shows maps of mass-weighted average `line-of-sight' (LOS, defined as the $+z$ direction in the simulation box) velocities for the three runs. While strong features, such as the filament at the lower-right, exist in all runs, much more complex velocity structure is present in the shock refinement runs. This is especially noticeable in the upper-right part of the maps, within the virial radius, where a near uniformly moving section of gas in the BaseRef run becomes much more dynamically complex with increased resolution.

Maps of mass-weighted average radial velocity are shown in the second row of Fig.~\ref{plt:TurbMaps}. Within the feedback generated outflow (shown in orange) the average outflow velocity is largely uniform across a large range of angles in the BaseRef run. In both shock refinement runs there is much more variation within the outflow, again implying considerably more complex dynamics. Even going from the ShockRef128 run to the ShockRef512 run there are clear differences, including many small, dense inflowing clouds within the CGM (that appear to be similar to those found in recent work by \citet{Nelson2020}). The filament inflowing from the right illustrates very well the systematic differences between the runs, as the inflow velocity is consistently higher throughout in the shock refinement runs compared to the BaseRef run. By comparing the first and second rows we also find that the inflow velocities of filaments are significantly higher than implied by the LOS velocities, with, for example, the lower-right filament having a LOS velocity of $\sim200$ km s$^{-1}$, but an actual inflow velocity of $\sim500$~km s$^{-1}$. This indicates how projection effects can significantly affect estimates of filamentary dynamics from observations.  

The increase in radial velocities corresponds to a higher average inflowing mass flux of cool-dense gas in the CGM of at least $20$ per cent, due to more mass condensing into denser and faster filaments. This is described further in Section \ref{Section:EnergyBudget} where we find a significant boost in the kinetic energy of the cool-dense gas phase at higher resolution. The increase in mass flux we find could be at least partially due to hot gas in the CGM becoming entrained into inflowing filaments through hydrodynamic instabilities \citep{Gronke2018, Gronke2020, Mandelker2020a, Mandelker2020b}. We note however that the gas filaments in our shock refinement simulations tend to already be denser and moving slightly faster \textit{outside} the accretion shock, where the mechanisms described in these works would not be as effective. However these effects may further enhance the cool-dense filaments once they enter the CGM. Further into the halo inflows are more disrupted by orbiting substructures as well as outflows, so we cannot easily disentangle the effect of resolution alone on the mass flux in these central regions.

\begin{figure} 
\centering
\includegraphics[width=1\linewidth]{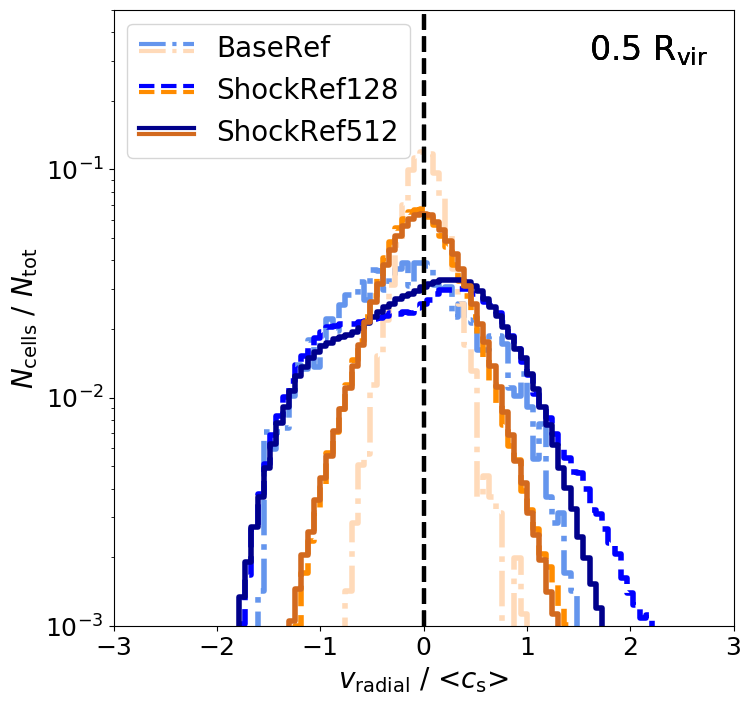}
\includegraphics[width=1\linewidth]{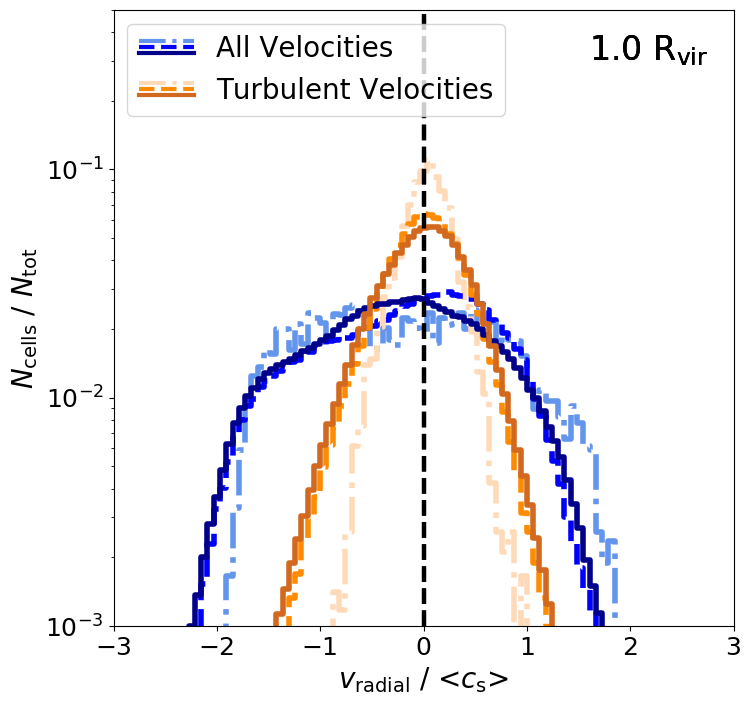}
\caption{Distribution of the radial components of gas velocities at $0.5 R_\mathrm{vir}$ (top panel) and at $R_\mathrm{vir}$ (bottom panel). The bulk velocities are shown in shades of blue and the turbulent velocities are in shades of orange. Only gas cells in the hot halo that are within a shell of thickness $\sim7$~kpc around each radius are included. Velocities are normalised to the mass-weighted average sound speed within each radial shell. The BaseRef, ShockRef128 and ShockRef512 runs are shown with dot-dashed, dashed and solid lines, respectively. The symmetric structure of turbulent velocities imply the halo exhibits turbulent pressure support. With increased refinement, we find a broader distribution of turbulent velocities, with larger extreme values.}
\label{plt:RadVelProf}
\end{figure}

\begin{figure} 
\centering
\includegraphics[width=1\linewidth]{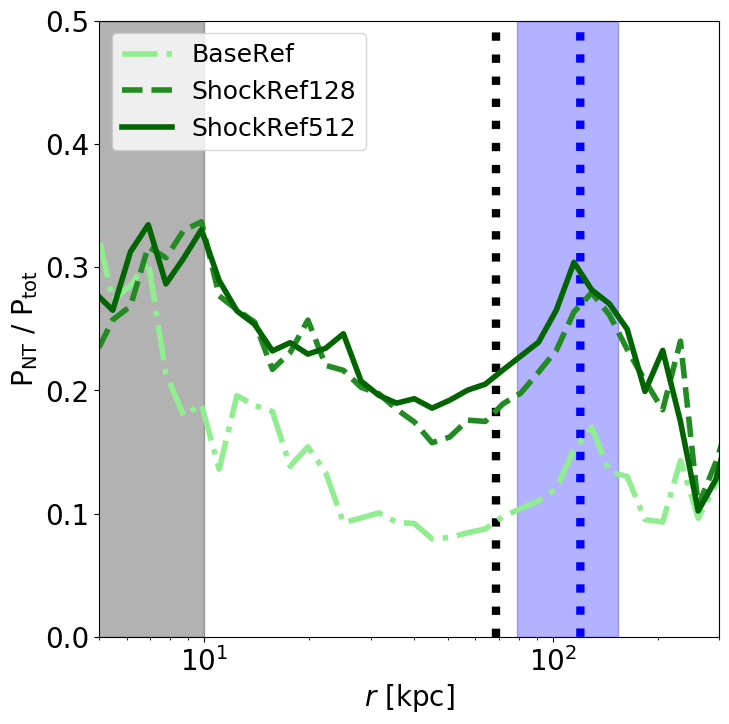}
\caption{Radial profiles of non-thermal pressure support fraction, as calculated in Equation \ref{eqn:ntpressure}, in the hot phase of the gaseous halo (see main text for the definition), comparing the three different simulations. The black, dotted vertical line indicates $R_\mathrm{vir}$ and the blue, dotted vertical line and shaded region indicate the median position and radial extent of the accretion shock in the ShockRef128 run. The shaded grey region denotes spatial scales below the minimum resolved turbulence scale of our method, as described further in the main text.}
\label{plt:NTPressure}
\end{figure}

As we have shown, shocks are also better resolved and more prevalent within our shock refinement simulations and, as astrophysical shocks are generally curved, we would naturally expect this to generate more vorticity and turbulent motions. Vorticity, $\mathbf{w} = \nabla \times \mathbf{u}$, is straightforward to calculate from the simulation output. However, it is not readily comparable to observations, which rely on measurements of velocities and velocity dispersions. We therefore present maps of LOS velocity dispersion in the third row of Fig.~\ref{plt:TurbMaps}, calculated relative to the mean LOS velocity. There are much more complex structures in the shock refinement velocity dispersion maps, with significant boosts in the wake of the accretion shocks, within filaments, and a significant increase within the virial radius\footnote{While not shown here, we have verified that gas vorticity follows similar qualitative trends to velocity dispersion.}.

Another way to quantify changes in turbulent motion is to separate velocities into turbulent and bulk components, which are treated in a way that mocks X-ray observations of galaxy clusters. Specifically, we first calculate turbulent velocities for individual cells using an adapted version of the multiscale filter method used in \citet{BourneJet}, based on the method described in \citet{VazzaTurbulence}. We can write the total velocity of each cell as 
\begin{equation} \label{Eq: VelComponents}
    \mathbf{v}_\mathrm{tot} = \mathbf{v}_\mathrm{bulk} + \mathbf{v}_\mathrm{turb}\,,
\end{equation}
where $\mathbf{v}_\mathrm{bulk}$ is the local bulk velocity and $\mathbf{v}_\mathrm{turb}$ is a turbulent velocity component. The local bulk velocity of a cell is first calculated as a mass-weighted average over a minimum number of nearest neighbour cells. This is then subtracted from the total velocity of a cell to give a turbulent velocity estimate. The number of neighbours is then iteratively increased until the turbulent velocity converges, within a tolerance factor. An additional complication we have to consider here is the varying spatial resolution between our runs, as they will naturally probe considerably different turbulent eddy sizes. To account for this effect we have therefore set the minimum number of initial neighbours to be inversely proportional to the mass of each cell, i.e. cells with two times the base target mass ($m_\mathrm{max} = 3.28\times 10^7$ M$_\odot$) are set to have $16$ neighbours and every other cell in both runs then starts the neighbour iteration with 
\begin{equation}
    N_\mathrm{NGB} = 16\frac{m_\mathrm{max}}{m_\mathrm{cell}}\,.
\end{equation}
This allows us to more directly compare turbulent velocities at a similar spatial scale. They are therefore calculated for eddy sizes approximately equal to the \textit{smallest spatial scale that the BaseRef run can reliably resolve}, where turbulent velocities are estimated across at least 16 neighbours. This corresponds to a physical scale of $\sim 10$~kpc. In this work we do not investigate the properties of smaller scale eddies in the shock refinement runs.

The bottom row of Fig.~\ref{plt:TurbMaps} shows maps of mass-weighted average turbulent velocity obtained with the method described above for the three runs. In the maps a marked increase in turbulent motion across the halo is evident in the shock refinement runs, with the magnitude particularly boosted in the central regions and in the wake of the accretion shocks. This corresponds well with our maps of velocity dispersion shown in the third row. Furthermore, we confirm similar qualitative trends in our basic Reference runs, albeit with more extreme turbulent velocities due to faster inflows and stronger accretion shocks. We note that this figure removes eEoS gas but still contains cold gas, to make the maps more comparable to the other velocity maps in Fig.~\ref{plt:TurbMaps}. Within the inflowing filaments and substructures, cold dense gas skews the turbulent velocities to lower values, so these regions should be treated with caution. For all other results using turbulent velocities we only consider hot halo gas, described below, though we note that the qualitative trends we find are the same regardless of the temperature-density cut applied.

To more faithfully compare our quantitative results with future X-ray observations, we also need to remove any cold clumps. This will also minimise the impact of eEoS gas, whose temperature at a given density is not realistic. To do this, we only include cells selected by a rescaled version of the method in \citet{RasiaCut}, as described in \citet{FABLE1}, henceforth referred to as hot halo gas. \citet{RasiaCut} identified a cooling phase of gas in their simulated clusters that satisfied the condition 
\begin{equation}
    T < N \times \rho^{0.25}\,,
\end{equation}
where $T$ and $\rho$ are the temperature and density of the gas and $N$ is a normalisation factor. This comes from the polytropic equation for an ideal gas, $T \propto \rho^{\gamma-1}$, assuming a polytropic index of $\gamma = 1.25$. \citet{RasiaCut} assumes a fixed normalisation factor $N = 3\times10^6$ keV cm$^{3/4}$ g$^{-1/4}$, where temperature is in keV and density is in g cm$^{-3}$. \citet{FABLE1} then rescaled this relation by the virial temperature of their haloes, $T_{500} \propto M_{500} / R_{500}$, relative to the mean virial temperature of the haloes in \citet{RasiaCut}. We note that these relations were found at $z=0$, when haloes can have a significantly different thermodynamic structure to $z=6$. However we have verified that performing the same mass rescaling as \citet{FABLE1} still gives a reasonable separation of the hot phase from the cooling components of the halo. 

Fig.~\ref{plt:RadVelProf} shows the distribution of radial turbulent and bulk velocities of cells in the hot gaseous halo within $7$~kpc annuli around $0.5 R_\mathrm{vir}$ and $R_\mathrm{vir}$. At both radii, though more noticeably at $R_\mathrm{vir}$, there is a significant inflowing component evident in the bulk velocities, even in the hot phase of the halo. There is also a tail at high outflow velocities, corresponding to recent bursts of feedback. In contrast, the turbulent velocity distributions are largely symmetric at both radii plotted. This therefore points to turbulence providing a source of non-thermal pressure support throughout these regions of the gaseous halo. Despite different halo masses and a very different redshift, qualitatively our results are similar to those found by \citet{VazzaTurbulence2}. We note that the impact of the shock refinement is to broaden the turbulent velocity distribution, reaching considerably higher extrema values, while still maintaining the symmetry indicative of turbulent pressure support. Moreover, the two shock refinement runs have largely similar turbulent velocity distributions, suggesting our calculations of turbulent velocity are starting to converge at the minimum resolved length scale in the BaseRef run (however we note that the average temperature of the halo is lower and there is less mass, and therefore less thermal energy, in the hot gas phase in the ShockRef512 run compared to the ShockRef128 run due to a more multiphase CGM).

With these differences in mind, we now estimate the fraction of non-thermal pressure support within the halo. We define this in the same way as \citet{VazzaTurbulence2}:
\begin{equation} \label{eqn:ntpressure}
    \frac{P_\mathrm{NT}}{P_\mathrm{tot}} = \frac{\rho v_\mathrm{turb}^2 / \alpha_\mathrm{r}}{\rho v_\mathrm{turb}^2 / \alpha_\mathrm{r} + \rho k_\mathrm{B} T / \mu_\mathrm{e} m_\mathrm{p}}\,,
\end{equation}
where $\rho$ is the gas density, $\mu$ is the mean molecular mass per electron, $T$ is the gas temperature and $\alpha_\mathrm{r} = 1$ if we consider only the radial velocity component and $3$ if we assume an isotropic dispersion. 

In Fig.~\ref{plt:NTPressure} we plot radial profiles of this non-thermal pressure ratio assuming an isotropic dispersion. We find a fairly consistent boost by a factor of $2$ in this fraction throughout the halo when the shock refinement scheme is active. The halo in each run approximately virialises at the same radius, meaning the bulk velocities of the hot halo gas are largely similar. This therefore indicates that it is the increase in turbulent velocity that changes in accord with Fig.~\ref{plt:TurbMaps}, leading to a sustained boost in turbulent pressure support out to and beyond the virial radius. This value is significantly higher than that found by \citet{VazzaTurbulence2}, however at $z=6$ this is to be expected due to more numerous inflows stirring the gaseous halo in this large overdensity than in similar mass haloes at lower redshifts. Deducing the non-thermal pressure support of galaxy clusters has important implications for calibrating biases in hydrostatic mass estimates, and our results suggest that these biases may be even more significant at high redshifts, which future X-ray missions such as \textit{ATHENA} will be able to probe.

\subsection{CGM Energy Budget} \label{Section:EnergyBudget}
\begin{figure} 
\centering
\includegraphics[width=1\linewidth]{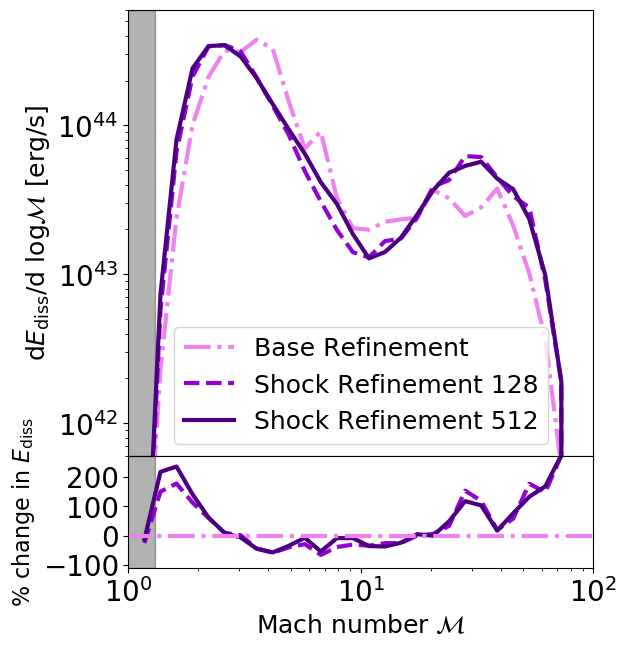}
\caption{\textit{Top panel}: distribution of energy dissipation rate as a function of Mach number for the BaseRef (dot-dashed), ShockRef128 (dashed) and ShockRef512 (solid) runs. \textit{Bottom panel}: percentage change in energy dissipation rate when the shock refinement scheme is active, as a function of Mach number. The grey area corresponds to $\mathcal{M} < 1.3$, where shocks are potentially spurious.}
\label{plt:EdissMach}
\end{figure}

\begin{figure} 
\centering
\includegraphics[width=\linewidth]{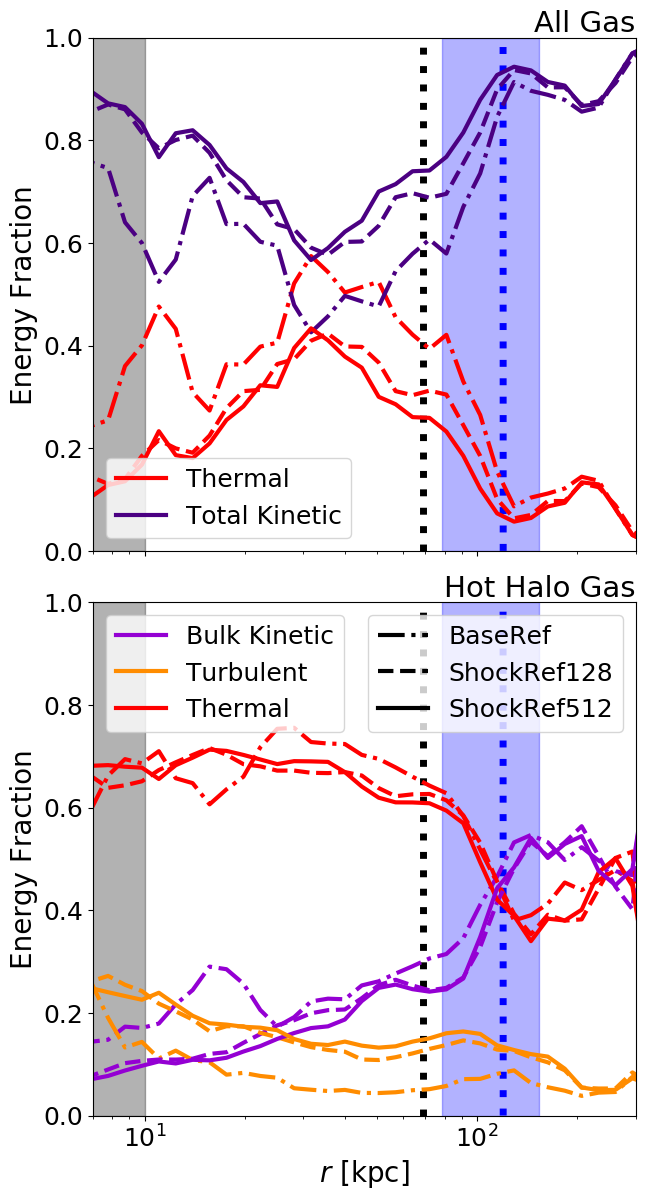}

\caption{Radial profiles of the \textit{energy fraction} stored in different forms within the gaseous halo for the 3 different refinement runs. \textit{Top panel}: Profiles including \textit{all} gas, showing the comparative fraction of energy in kinetic and thermal components. \textit{Bottom panel}: Profiles for hot halo gas only, further separating bulk kinetic from turbulent motion. For both panels, the black dotted vertical line represents $R_\mathrm{vir}$, and the shaded grey area indicates the region where turbulent velocities are unresolved (see main text for details). The blue dotted line and shaded region indicates the median position and radial extent of the accretion shock in the ShockRef128 run, respectively.}
\label{plt:EnergyFractions}
\end{figure}

\begin{figure} 
\centering
\includegraphics[width=\linewidth]{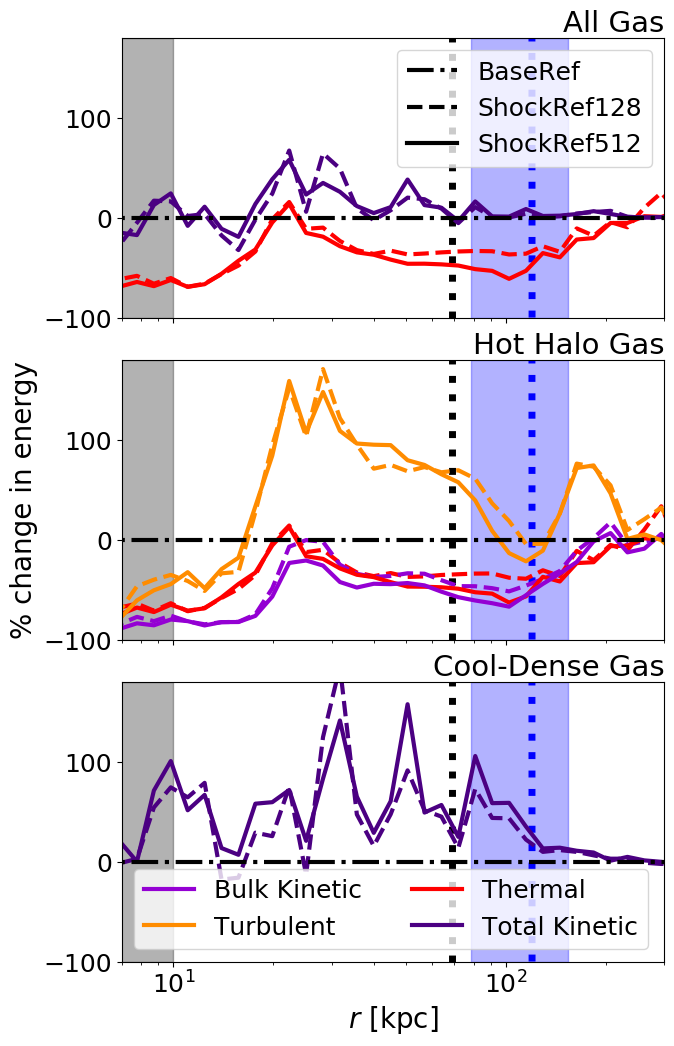}

\caption{Radial profiles of the percentage change in \textit{absolute energy} stored in different components with respect to the BaseRef run. \textit{Top panel}: Profiles including \textit{all} gas, showing the change in total energy in kinetic and thermal components. \textit{Centre panel}: Profiles for hot halo gas only, showing a large increase in turbulent energy and a suppression in bulk kinetic and thermal energy. \textit{Bottom panel}: Profiles for cool-dense gas, defined in the same way as Fig. \ref{plt:ColdGasFrac}, showing an increase in kinetic energy. For all panels, the black dotted vertical line represents $R_\mathrm{vir}$, and the shaded grey area indicates the region where turbulent velocities are unresolved (see main text for details). The blue dotted line and shaded region indicates the median position and radial extent of the accretion shock in the ShockRef128 run, respectively. We note that the gravitational potential energy does not appreciably vary between the runs, given that the total mass distribution of the halo does not significantly change.}
\label{plt:EnergyChanges}
\end{figure}

The variations in the topology of shocks and the changes in cooling rates, along with the different distribution of gas in different phases with shock refinement active, will all clearly have an impact on the energy budget of the halo.  

To examine this we first look at how energy dissipation in the shocks themselves varies with refinement scheme in Fig.~\ref{plt:EdissMach}, where the energy dissipation rate as a function of Mach number is shown. In the ShockRef128 run we find a boost of nearly $\sim200$ per cent in energy dissipation at low Mach numbers, where the largest peak of the distribution shifts towards lower Mach number values. When we increase the resolution even further in the ShockRef512 run, this peak becomes somewhat higher. These weak shocks often corresponds to regions around the edge of filaments and substructures, shown in red and dark blue in the bottom panels of Fig.~\ref{plt:ClusterVis}. Here, post-shock gas generally has a high density but still a fairly low temperature due to the weakness of the shock, meaning the gas is still close to the peak of the hydrogen cooling curve. Any transfer from kinetic to thermal energy in these locations is therefore efficiently radiated away, as indicated in the third row of Fig.~\ref{plt:ClusterVis} where we find increased cooling rates in and around inflowing filaments which ultimately leads to a lower thermal energy of the gaseous halo with increased resolution.

At high Mach numbers, corresponding to yellow-white parts of the bottom panels of Fig.~\ref{plt:ClusterVis}, we also find a boost in energy dissipation. This is largely the same between the two shock refined runs, suggesting the energy dissipation rate of strong shocks is beginning to converge. Given the vastly higher number of cells in the shock refinement runs, we naturally find many more shocked cells within the halo. However, the total \textit{mass} of shocked cells in these haloes is considerably lower, as shock widths become much narrower due to less numerical broadening (see bottom panels of Fig.~\ref{plt:ClusterVis}). 

Fig.~\ref{plt:EnergyFractions} shows radial profiles of the gas \textit{energy fraction} stored in thermal and kinetic components. In the top panel we show these for \textit{all} gas. The thermal fraction increases significantly within the accretion shock (vertical, dashed blue line) as infalling gas is shocked and thermalised, though we note the peak thermal fraction is reduced in the shock refined runs due to the development of a more multiphase CGM.
In the bottom panel, we show radial profiles of energy fraction in the hot phase of the gaseous halo. Here we further separate the total kinetic energy into bulk kinetic and turbulent components by separating the velocity components as described in Equation~\ref{Eq: VelComponents}. The most notable result here is the amount of energy stored in turbulent motions in the CGM, increasing from $4$ per cent in the BaseRef run to $\sim15$ per cent in the ShockRef128 run and slightly higher values than that in the ShockRef512 run. This comes at the expense of both the thermal and bulk kinetic energy of the hot phase. We do not show equivalent energy fraction profiles for the cold-dense phase, as this is overwhelmingly dominated by kinetic energy.

In Fig. \ref{plt:EnergyChanges} we show the percentage change in \textit{absolute energy} within each component with respect to the BaseRef run. The top panel shows this for \textit{all} gas, where we find a suppression in thermal energy throughout the CGM in both shock refinement runs (by up to $\sim30$ and $\sim50$ per cent in the ShockRef128 and ShockRef512 runs, respectively). This is in line with our earlier finding of an increase in cold gas penetrating the hot halo at higher resolution, as well as a higher cooling rate driven by larger energy dissipation. The total kinetic energy of cold gas is higher inside the accretion shock, as more gas is associated with the quickly inflowing filaments, though as we discuss below this varies between gas phases. The central panel of Fig. \ref{plt:EnergyChanges} plots the energy changes in the hot halo gas, showing a large boost in turbulent energy of more than $\sim 100$ per cent in both shock refinement runs, coming at the expense of both the thermal and bulk kinetic components. This is partly due to better defined shock surfaces in the gaseous halo, most of which are curved, generating more vorticity and turbulent motions as we discussed in Section~\ref{Section:Turbulence}. In boosting the resolution we also capture more of the turbulent cascade, meaning more energy is stored for longer in turbulent motions before eventually being thermalised. We find that the bulk kinetic energy of cool-dense gas is increased by a significant amount within the accretion shock in both shock refinement runs, shown in the bottom panel of Fig.~\ref{plt:EnergyChanges}, regularly exceeding the BaseRef values by more than $50$ per cent. This is driven by an increase in mass in the cool-dense phase as discussed in Section~\ref{Section:ColdGas} which is predominantly part of denser, faster inflowing filaments. This is also indicative of the increase in inflowing mass flux previously discussed.

\subsection{Dense, Neutral Gas and Star Formation} \label{Section:StarFormation}
\begin{figure*} 
\centering
\includegraphics[width=\linewidth]{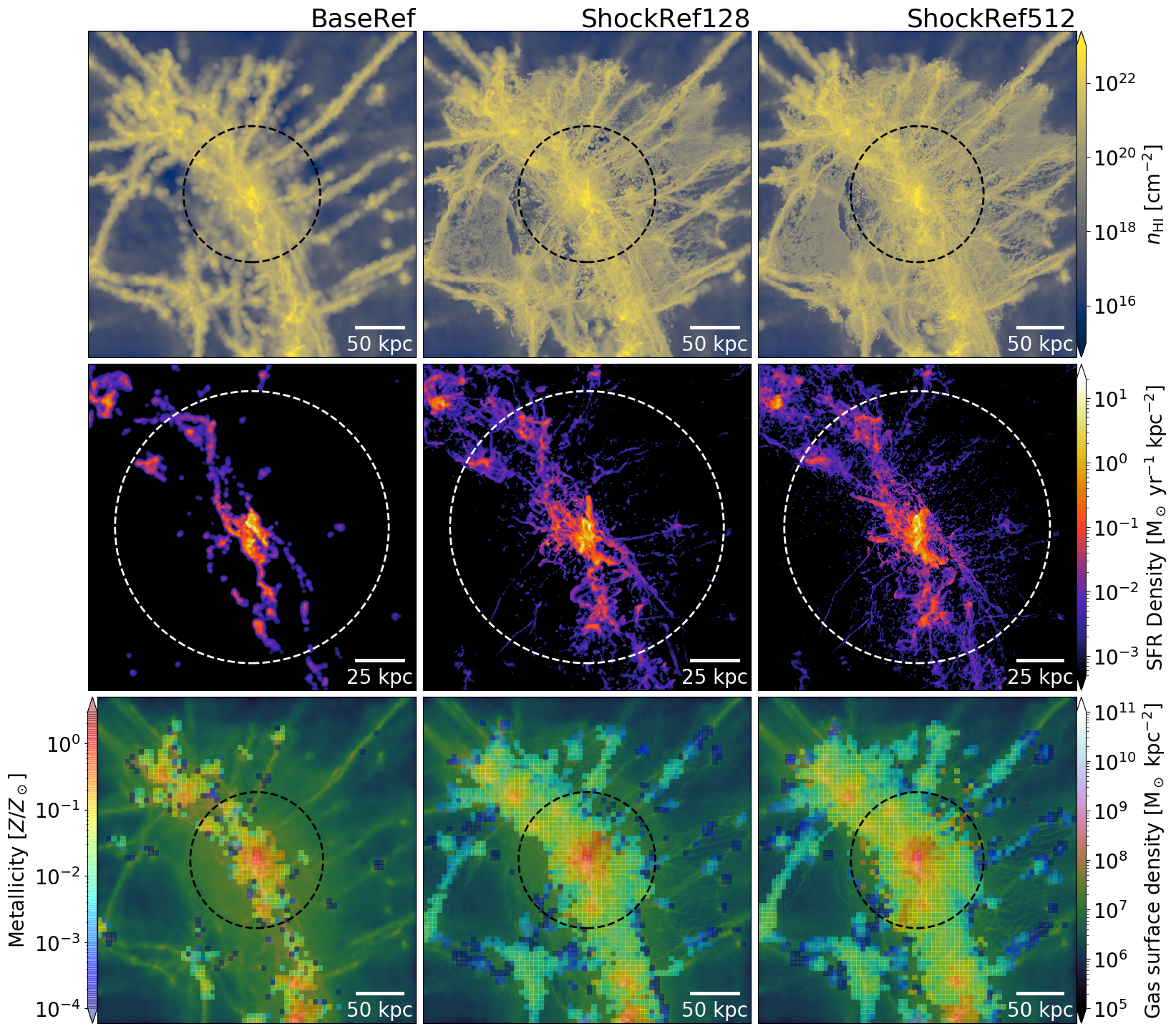}
\caption{\textit{From top to bottom}: maps of H\textsc{i} surface density, star formation rate surface density and gas surface density, overlaid with a map of star particle locations coloured by mass-weighted average stellar metallicity. The BaseRef run is shown in the left-hand column, the ShockRef128 run is in the central column and the ShockRef512 is shown in the right-hand column. All quantities are projected over a slice of thickness $2\,R_\mathrm{vir}$, and black/white dashed circles indicate $R_\mathrm{vir}$. There are considerably more areas within the CGM at larger H\textsc{i} column densities with higher resolution. Star formation (and stars) are much more spread out through the halo, particularly within inflowing filaments where we find the formation of very metal-poor stars.}
\label{plt:SFRMaps}
\end{figure*}

\begin{table}
	\centering
	\caption{H\textsc{i} covering fractions within $R_\mathrm{vir}$ and $2\,R_\mathrm{vir}$ for four different column density thresholds, comparing the three different refinement runs. We note that the absolute covering fraction values at $z=6$ will be sensitive to the implementation of the UV background, but we expect the changes due to resolution to persist. Values are calculated over a slice of depth $2\,R_\mathrm{vir}$.}
	\label{tab:coverfrac}
	\begin{tabular}{c|c|c|c}
        \hline
        Threshold & Run & $R_\mathrm{vir}$ & $2\,R_\mathrm{vir}$ \\
        \hline
        \hline
        \multirow{3}{*}{$\log (n_\mathrm{H\textsc{i}} / \mathrm{cm}^{-2}) > 17$} & BaseRef & 0.985 & 0.910 \\
        & ShockRef128 & 0.996 & 0.997 \\
        & ShockRef512 & 0.999 & 0.998 \\
        \hline
        \multirow{3}{*}{$\log (n_\mathrm{H\textsc{i}} / \mathrm{cm}^{-2}) > 18$} & BaseRef & 0.944 & 0.721 \\
        & ShockRef128 & 0.962 & 0.941 \\
        & ShockRef512 & 0.983 & 0.969 \\
        \hline
        \multirow{3}{*}{$\log (n_\mathrm{H\textsc{i}} / \mathrm{cm}^{-2}) > 19$} & BaseRef & 0.835 & 0.592 \\
        & ShockRef128 & 0.903 & 0.828 \\
        & ShockRef512 & 0.950 & 0.908 \\
        \hline
        \multirow{3}{*}{$\log (n_\mathrm{H\textsc{i}} / \mathrm{cm}^{-2}) > 20$} & BaseRef & 0.621 & 0.417 \\
        & ShockRef128 & 0.747 & 0.586 \\
        & ShockRef512 & 0.832 & 0.651 \\
        \hline
	\end{tabular}
\end{table}

We have shown how boosting resolution increases the amount of cool-dense gas in the gaseous halo, so we would therefore expect an increase in the amount of neutral H\textsc{i} gas as well. This provides a reservoir of fuel for star formation, thereby influencing the number of stars that form as well as where they are located.

We first investigate how the neutral H\textsc{i} gas covering fraction changes between the three runs. To do this we calculate the H\textsc{i} gas content of each cell using the prescription of \citet{BirdHIConversion}. We show maps of H\textsc{i} column density in the top row of Fig.~\ref{plt:SFRMaps}, where we find significant increases in the H\textsc{i} column density especially at radii outside the virial radius (shown as a black dashed circle). In Table~\ref{tab:coverfrac} we list H\textsc{i} covering fractions within $R_\mathrm{vir}$ and $2\,R_\mathrm{vir}$ for the three different refinement runs. For all four column density thresholds we find an increase in the covering fraction within both radii, with the most striking increases occurring at the highest column densities and at larger radii. Visually this corresponds to yellow-coloured regions in the top row of Fig.~\ref{plt:SFRMaps} with column densities above $10^{19}$cm$^{-2}$. This is especially true in-between filamentary structures such as, for example, on the centre-left and upper-right parts of the maps. This appears similar to the granular morphology found in these regions, arising from the shattering of cosmic sheets through thermal instabilities, described in \citet{Mandelker2019}. We note that as we increase the resolution to that of the ShockRef512 run the neutral H\textsc{i} covering fractions keep increasing, especially at high column densities, showing only weak convergence. This was also found in both the IGM and the CGM by \citet{Mandelker2019} and \citet{vdVoortCGM}, respectively.

Qualitatively our findings agree with those of \citet{HummelsCGM}, \citet{Mandelker2019} and \citet{vdVoortCGM}, though detailed comparisons are not possible due to the high redshift of our simulations. \citet{SureshCGM} do not find a resolution dependence for H\textsc{i} covering fractions, though as previously discussed their base solution already has very high resolution beyond that of even the ShockRef512 so this is perhaps to be expected. Our results suggest that simulated H\textsc{i} covering fractions do depend significantly on resolution (and are not converged yet at the highest resolution explored here), which will clearly have implications for the prediction of the distribution of Damped Lyman-$\alpha$ and Lyman-limit systems (DLAs and LLSs, respectively) as well as observations of the Lyman-$\alpha$ forest. We note that the absolute values for covering fractions we find at $z=6$ will be very sensitive to our implemented UV background model, but the changes we find with respect to resolution are expected to persist for different reionisation histories and at lower redshifts as well. 

Overall by $z=6$ in the ShockRef128 run we find an increase of $\sim20$ per cent in the stellar mass within the halo's half-mass radius, and the same percentage increase within the virial radius compared to the BaseRef run. In the ShockRef512 run this becomes a $\sim30$ per cent increase.  We explain this increase by looking at the central panels of Fig.~\ref{plt:SFRMaps}, which show maps of star formation rate surface density for the three runs. Star formation is more widespread throughout the halo, extending further out from the central galaxy than in the BaseRef run. Star formation also occurs within filaments within the CGM, which are predominantly made up of cold, low-metallicity gas. 

In the bottom panels of Fig.~\ref{plt:SFRMaps} we overlay a gas density map with the distribution of stellar metallicities. While the metallicities in the central galaxy are largely similar in the three runs, reaching super-Solar levels in the innermost regions, there is a more continuous gradient of decreasing metallicity to larger radii in the shock refinement runs. In several places very low-metallicity stars, with $Z/Z_\odot < 10^{-3}$, have formed in the inflowing filaments that are then fed into the halo, implying cosmic filaments could be a non-negligible source of metal poor stars\footnote{We note here we do not self-consistently model the formation of Pop III stars or have a metallicity floor at very high redshift, so very low absolute values of stellar metallicity should be treated with caution. However, previous work that has explicitly modelled Pop III stars have found that they should not lead to metallicities above $10^{-3}$ $Z_\odot$ in filaments \citep{Wise2012}, implying our results are reasonable.}. These results qualitatively agree with \citet{Mandelker2018}, who predict that stars and globular clusters could form in cold, inflowing filaments at high redshift.

We caution that our simulations employ a fairly simplistic sub-grid star formation model based on the gas density threshold, so the increased densities we find with boosted resolution will naturally increase star formation. Nevertheless, the predictions made in this work could provide useful tests of star formation models with the next generation of observations from the James Webb Space Telescope (\textit{JWST}). Using deep NIRCam exposures we would expect regions with star formation surface density of $0.3$ M$_\odot$ yr$^{-1}$ kpc$^2$ and above to be detectable even at $z=6$, however detecting nebular emission from these regions will likely be much more difficult (R. Maiolino, priv. comm.).

\section{Conclusions} \label{Section:Conclusion}

In this work we designed a novel `shock refinement' scheme to target increases in numerical resolution around shocks on-the-fly in hydrodynamical galaxy formation simulations. We verified the accuracy and effectiveness of the scheme by comparing simple 2D Sedov-Taylor blast wave simulations against known analytic solutions. The new scheme gives us a very flexible and physically motivated way to boost resolution within interesting regions of galactic haloes, such as at the boundary of the CGM and IGM, without the significant increase in simulation runtime that would come from a global increase in resolution.

As a proof-of-concept we then applied our scheme to cosmological zoom-in simulations of a very rare overdensity containing a $\sim3\times 10^{12}$~M$_\odot$ halo at $z=6$, where we compared the effects of the base refinement scheme (BaseRef) with shock refinement runs with mass resolution increased by $128$ and $512$ (ShockRef128 and ShockRef512, respectively). We performed this analysis for both a basic physical model, with only primordial cooling, and the full FABLE model and found that our results are robust against changes in the modelling of galaxy formation physics. The threshold Mach number, above which shocked cells and their neighbours refine, was chosen in this work to be $\mathcal{M} = 1.5$ as we targeted boosted resolution around both weak shocks around cosmic filaments and strong halo accretion shocks. 

When the shock refinement scheme is activated we find a significant boost in the number of cool, dense and metal-poor structures within the hot gaseous halo. This gives rise to a much more multiphase CGM more in line with recent observations, which are still albeit at a much lower redshift. These structures predominantly correspond to primordial filaments passing through the accretion shock without being significantly shock heated. We demonstrated this using a novel visualisation technique, where we defined a closed 3D surface corresponding to the accretion shock. Combining this with the filament finding software DisPerSE showed that filaments tend to penetrate the accretion shock where the local Mach number is very low. The better refined and more numerous filaments in the shock refined simulations also have higher cooling rates, leading to a $\sim30$ and $\sim50$ per cent increase in cool-dense gas within the accretion shock, for the ShockRef128 and ShockRef512 runs, respectively, largely at the expense of the warm-dense phase. 

Higher resolution in the halo allows us to better capture small-scale motions in the shock refinement run, leading to a much more complex kinematic structure. The well-resolved shocks within the halo are predominantly curved, leading to a significant boost in velocity dispersion and turbulent motions within the gaseous halo, especially in the wake of strong shocks. We find the average turbulent velocity increases significantly throughout the halo, corresponding to a wider distribution of turbulent velocities with larger extreme values. This boosted turbulent motion acts as a source of non-thermal pressure support, which we find corresponds to $20 - 30$ per cent of the total pressure support in the hot halo component of both shock refinement runs compared to $10-15$ per cent in the base refinement run. 

We find that energy dissipation is boosted in shocks, particularly for very weak and very strong shocks, where the dissipation rate is increased by $>100$ per cent in both shock refinement runs. Gas cooling rates are also increased in the CGM, caused by better resolving denser filaments when shock refinement is active. Both of these factors lead to changes in the energy budget of the halo. There is a suppression in the thermal energy of the halo of up to $30$ per cent in the ShockRef128 run and up to $50$ per cent in the ShockRef512 run as the more multiphase CGM has a lower average temperature and more energy is radiated away at weak shocks. When we further separate kinetic energy into bulk and turbulent motions, for the hot halo gas, we find a boost of more than $100$ per cent in turbulent energy in the halo as we capture larger curved shock surfaces and more of the turbulent cascade. The total kinetic energy in cool-dense gas is enhanced by at least $20$ per cent throughout the CGM, and regularly exceeds $50$ per cent higher. This is aligned with an increase in inflowing mass flux, which for cool-dense gas shows similar increases.

Finally, we consider the impacts that the shock refinement scheme might have on observational predictions. We find that the covering fraction of neutral H\textsc{i} gas is significantly increased in the CGM with higher resolution, which has implications for predicting the distribution and redshift evolution of DLAs, LLSs and the Lyman-$\alpha$ forest. The increase in cool, dense gas at large radii allows star formation to proceed further out from the central galaxy in the shock refinement runs than in the BaseRef run. Stars are considerably more spread out in both shock refinement runs and have a broad range of metallicities, including very metal-poor stars forming along inflowing filaments that deep \textit{JWST} observations are likely to probe.  

The results we find in this work suggest that there are still numerical barriers that need to be overcome to piece together a fuller, more robust picture of galaxy formation. This will be especially true in large volume cosmological simulations, where the numerical resolution is necessarily worse, especially in the CGM and IGM, than the zoom-ins we study here. Our new shock refinement scheme provides a useful tool to address this, as it more efficiently boosts numerical resolution around regions of interest, like accretion shocks at the boundary of the CGM and IGM. Within this study we have pushed the Mach threshold of our scheme as low as possible, to quantify effects from boosting resolution around the majority of shocks, highlighting the importance of adequate numerical resolution in studying the CGM. In future, the versatility of our scheme means it can be readily extended and applied to different physical situations, including resolving individual supernova remnants or shocks in and around AGN jets in galaxy formation simulations. It could also be used to achieve zoom-in level resolution within the CGM of \textit{multiple} haloes within a simulation without globally increasing resolution to prohibitive levels.

\section*{Acknowledgements}

The authors would like to thank the anonymous referee and scientific editor for useful comments that improved the manuscript. They would also like to thank Nick Henden, Martin Bourne, Roberto Maiolino, Martin Haehnelt and Clotilde Laigle for useful discussions and assistance. JB and DS acknowledge support from the Science, Technology and Facilities Council (STFC) and the ERC Starting Grant 638707 `Black holes and their
host galaxies: co-evolution across cosmic time'. 

This work was performed using resources provided by: the Cambridge Service for Data Driven Discovery (CSD3) operated by the University of Cambridge Research Computing Service (www.csd3.cam.ac.uk), provided by Dell EMC and Intel using Tier-2 funding from the Engineering and Physical Sciences Research Council (capital grant EP/P020259/1); the DiRAC@Durham facility managed by the Institute for Computational Cosmology on behalf of the STFC DiRAC HPC Facility (www.dirac.ac.uk). The equipment was funded by BEIS capital funding via STFC capital grants ST/P002293/1, ST/R002371/1 and ST/S002502/1, Durham University and STFC operations grant ST/R000832/1. DiRAC is part of the National e-Infrastructure.

\section*{Data Availability}
The data used in this work may be shared on reasonable request to the authors.

%%%%%%%%%%%%%%%%%%%%%%%%%%%%%%%%%%%%%%%%%%%%%%%%%%



\bibliographystyle{mnras}
\bibliography{References}

%%%%%%%%%%%%%%%%%%%%%%%%%%%%%%%%%%%%%%%%%%%%%%%%%%

%%%%%%%%%%%%%%%%% APPENDICES %%%%%%%%%%%%%%%%%%%%%

%%%%%%%%%%%%%%%%%%%%%%%%%%%%%%%%%%%%%%%%%%%%%%%%%%

% Don't change these lines
\bsp	% typesetting comment
\label{lastpage}
\end{document}